\documentclass[journal]{IEEEtran}
\usepackage{bbm}
\usepackage{dsfont}
\usepackage{yfonts}
\usepackage{bm}
\usepackage{amsmath}
\usepackage{amssymb}
\usepackage{amsfonts}
\usepackage{mathrsfs} 
\usepackage{stfloats}
\usepackage{amsthm}
\usepackage{subfigure}
\usepackage{graphicx}
\usepackage[normalem]{ulem}
\usepackage{graphicx,amssymb,amsmath}
\usepackage{multicol}
\usepackage[noadjust]{cite}
\usepackage{setspace}
\usepackage{midfloat}
\usepackage[normal]{threeparttable}
\usepackage{amsthm}
\usepackage{cuted}
\usepackage{cases,subeqnarray}
\usepackage{bm,multirow,bigstrut}
\usepackage{textcomp}
\usepackage{latexsym,bm}
\usepackage{booktabs,changebar}
\usepackage{xcolor}
\usepackage{algorithmic}
\usepackage[ruled,vlined]{algorithm2e}
\usepackage{extarrows}
\usepackage{tablefootnote}
\newcommand{\RomanNum}[1]{\uppercase\expandafter{\romannumeral #1\relax}}
\ifCLASSOPTIONcompsoc
\usepackage[nocompress]{cite}
\else
\usepackage{cite}
\fi

\IEEEoverridecommandlockouts
\begin{document}

\title{Mixed-ADC/DAC Multipair Massive MIMO Relaying Systems: Performance Analysis and Power Optimization}

\author{Jiayi~Zhang,~\IEEEmembership{Member,~IEEE,}
        Linglong~Dai,~\IEEEmembership{Senior Member,~IEEE,}
        Ziyan~He,\\
        Bo~Ai,~\IEEEmembership{Senior Member,~IEEE,}
        and~Octavia A. Dobre,~\IEEEmembership{Senior Member,~IEEE}

\thanks{This work was supported in part by the National Natural Science Foundation of China (Nos. 61601020 and 61725101), the Beijing Natural Science Foundation (Nos. 4182049 and L171005), the open research fund of National Mobile Communications Research Laboratory, Southeast University (No. 2018D04), Key Laboratory of Optical Communication and Networks (No. KLOCN2018002), National Key Research and Development Program (No. 2016YFE0200900), Major projects of Beijing Municipal Science and Technology Commission (No. Z181100003218010), National key research and development program under (No. 2016YFE0200900), and the Natural Sciences and Engineering Research Council of Canada (NSERC) through its Discovery program.}%
\thanks{J. Zhang is with the School of Electronic and Information Engineering, Beijing Jiaotong University, Beijing 100044, P. R. China. He is also with National Mobile Communications Research Laboratory, Southeast University, Nanjing 210096, P. R. China (e-mail: jiayizhang@bjtu.edu.cn).}
\thanks{L. Dai is with Department of Electronic Engineering, Tsinghua University, Beijing 100084, P. R. China (e-mail: daill@tsinghua.edu.cn).}
\thanks{Z. He was with Department of Electronic Engineering, Tsinghua University, Beijing 100084, P. R. China. He is now with School of Electrical and Computer Engineering, Georgia Institute of Technology, Atlanta, GA 30332 USA (e-mail: heziyan@gatech.edu).}
\thanks{B. Ai is with the State Key Laboratory of Rail Traffic Control and Safety, Beijing Jiaotong University, Beijing 100044, China (e-mail: boai@bjtu.edu.cn). }
\thanks{O. A. Dobre is with the Faculty of Engineering and Applied Science, Memorial University, St. John's,
NL A1B 3X5, Canada (e-mail: odobre@mun.ca).}
}

\maketitle
\vspace{-1.75cm}
\begin{abstract}
High power consumption and expensive hardware are two bottlenecks for practical massive multiple-input multiple-output (mMIMO) systems. One promising solution is to employ low-resolution analog-to-digital converters (ADCs) and digital-to-analog converters (DACs). In this paper, we consider a general multipair mMIMO relaying system with a mixed-ADC/DAC architecture, in which some antennas are connected to low-resolution ADCs/DACs, while the rest of the antennas are connected to high-resolution ADCs/DACs. Leveraging on the additive quantization noise model, both exact and approximate closed-form expressions for the achievable rate are derived. It is shown that the achievable rate can approach the unquantized one by using only 2-3 bits of resolutions. Moreover, a power scaling law is presented to reveal that the transmit power can be scaled down inversely proportional to the number of antennas at the relay. We further propose an efficient power allocation scheme by solving a complementary geometric programming problem. In addition, a trade-off between the achievable rate and power consumption for different numbers of low-resolution ADCs/DACs is investigated by deriving the energy efficiency. Our results reveal that the large antenna array can be exploited to enable the mixed-ADC/DAC architecture, which significantly reduces the power consumption and hardware cost for practical mMIMO systems.
\end{abstract}

\vspace{-0.5cm}

\IEEEpeerreviewmaketitle

\section{Introduction}
As one of the disruptive technologies for {the} fifth-generation (5G) wireless communications, massive multiple-input multiple-output (mMIMO) has attracted extensive research interests in recent years \cite{wong2017key,marzetta2016fundamentals,yadav2018all,morgado2018survey}. By exploiting quasi-orthogonal random channel vectors between different users, mMIMO can mitigate the inter-user interference to provide high spectral efficiency and energy efficiency via simple linear signal processing, e.g., maximum-ratio combining (MRC) and zero-forcing (ZF) precoding. On the other hand, relaying is an important way of extending coverage and providing uniform service. The inter-user interference of the multiuser relaying system can be suppressed by equipping the relay with a large number of antennas \cite{zhang2017spectral,wang2017performance,jin2015ergodic}.

The practical implementation of an mMIMO relaying system with hundreds or even thousands of antennas is a significant challenge \cite{bjornson2017massive,zhang2018low,Esswie2017novel}. {{For example, the perfect synchronization is difficult in mMIMO relaying system. One possible solution is to employ rateless network coding \cite{li2010cooperative}}}. Typically, each antenna in mMIMO systems is connected to {an analog-to-digital converter (ADC) and a digital-to-analog converters (DAC)} in the radio frequency (RF) chain, respectively. It is well known that the power consumption and hardware cost of ADCs and DACs linearly increase with the bandwidth and exponentially increase with the number of quantization bits \cite{lee2008analog-to-digital}. Thus, high-resolution ADCs and DACs (e.g., 8-12 bits for commercial use) will result in high power consumption and hardware cost in practical mMIMO relaying systems.

%

To solve this challenging problem, a promising solution is to replace power-hungry high-resolution ADCs and DACs (e.g., 8-12 bits) with low power low-resolution ADCs and DACs (e.g., 1-3 bits) \cite{zhang2016spectral,fan2015uplink}. However, significant signal processing challenges and complex front-end designs (e.g., channel estimation, phase/frequency synchronization, and multiuser detection) inevitably occur due to the strong nonlinear characteristic of coarse quantization \cite{zhang2016achievable}. As recently reported in \cite{li2017channel}, a high signal-to-noise ratio (SNR) channel estimation error floor exists due to the one-bit quantization. {Furthermore,} the authors in \cite{liang2016mixed} proposed a mixed-ADC architecture, where only a small fraction of ADCs are high-resolution, to facilitate the aforementioned signal processing and the establishment of front-end designs. {{For example, the CSI for each antenna can be obtained by using the high-resolution ADCs in a round-robin manner, which has been clearly explained in \cite{liang2016mixed}.}} {{Moreover, the mixed-ADC architecture is economically beneficial and easier to implement compared with architectures with uniform converter resolution, as it adds some antennas with low-resolution ADCs to the existing high-resolution conventional MIMO system \cite{yuan2017distributed}}}.

\subsection{Related Works}
Most of recent works focused on the single-hop mMIMO system with a mixed-ADC architecture. For instance, the mutual information of mixed-ADC mMIMO systems has been investigated in \cite{liang2016mixed,liang2016frequency}, which reveals that the mixed-ADC architecture is able to approach the ideal channel capacity of unquantized systems over both frequency-flat \cite{liang2016mixed} and frequency-selective fading channels \cite{liang2016frequency}. In addition, the achievable rate performance of multi-user mMIMO systems with a mixed-ADC architecture is comparable for Rayleigh \cite{tan2016spectral} and Rician fading channels \cite{zhang2017performance}. This architecture can achieve a better energy-rate trade-off compared with the ideal infinite-resolution and low-resolution ADC architectures. By applying probabilistic Bayesian inference, a family of detectors for mixed-ADC mMIMO systems was developed in \cite{zhang2016mixed}. It proves that the available high-resolution ADCs are practically essential since they can effectively eliminate the error floor of a relaxed Bayesian detector. {{Given the energy constraint at the base station (BS), the sum achievable rate has been maximized in \cite{pirzadeh2018spectral}, which shows that the optimal rate can be obtained by using only one-bit ADCs in most realistic scenarios.}} Moreover, \cite{xu2017performance} considered the downlink mMIMO with both mixed-resolution DACs at the BS and mixed-resolution ADCs at the user side.
These important contributions have shown that the power consumption and hardware cost of the single-hop mMIMO system with a mixed-ADC architecture can be considerably reduced while keeping most of the gains in the achievable rate.

In contrast to single-hop systems, very little attention has been paid to the two-hop mMIMO relaying system with both mixed-resolution ADCs and mixed-resolution DACs. Very recently, the authors in {\cite{kong2017multipair,dong2017efficient,liu2017multiuser,kong2017full,jia2016optimal} investigated the performance of a multipair mMIMO relaying system with low-resolution ADCs and DACs at the relay.} The achievable rate of such a system is limited by using very coarse quantization (e.g., one-bit). In this paper, we consider a more general architecture, where ADCs and DACs with arbitrary resolution profile are employed at the relay to achieve a possibly higher rate.

\subsection{Contributions}
In this paper, we focus on a general two-hop \emph{mixed-ADC/DAC mMIMO relaying system, where some antennas are connected to low-resolution ADCs/DACs, while the rest of the antennas are connected to high-resolution ADCs/DACs}. This study aims to analyse the performance analysis of the multipair mMIMO relaying system with arbitrary quantization noise, which is in contrast to the previous study \cite{kong2017multipair} that employs only one-bit ADCs and DACs. We demonstrate that the achievable rate of the considered system can approach that of the ideal unquantized system. The main contributions of this paper are summarized as follows:

\begin{itemize}
\item Leveraging on AQNM, we present a unified framework to derive the exact closed-form expressions for the achievable rate of mixed-ADC/DAC mMIMO systems. {{Compared with the Bussgang theorem used in \cite{kong2017multipair}, the AQNM can offer analytical tractability for multi-bit ADCs and DACs.}} Furthermore, approximate achievable rate expressions are derived by using asymptotic arguments. These results can provide insights into the effects of the number of relay antennas, user transmit power, quantization bits, and the fraction of high-resolution quantizers on the achievable rate, respectively.
\item The power-scaling law of the mixed-ADC/DAC architecture is investigated for power saving in the data transmission phase of the relay. Our results reveal that when the number of relay antennas, $M$, gets asymptotically large, the transmit power of each antenna can be scaled down by $1/M$ without any rate loss for the considered system. Moreover, the achievable rate gap between the mixed-ADC/DAC relay system and the unquantized system is a constant in the low power regime.
\item In order to compensate for the rate degradation caused by the coarse quantization, a low-complexity power allocation algorithm is proposed for the considered system. The power allocation problem can be solved by transforming it into a sequence of geometric programming (GP) problems.
\item Finally, using a generic power consumption model, we study the effects of the fraction of high-resolution ADCs and quantization bits on energy efficiency. In order to maximize energy efficiency, {{the optimal number of quantization bits is derived through numerical computations.}} Furthermore, our analysis proves that the considered system can significantly reduce power consumption and hardware cost while maintaining considerable performance.
\end{itemize}

\subsection{Outline}
The remaining parts of the paper are structured as follows. The mixed-ADC/DAC multipair mMIMO relaying system model is briefly introduced in Section \ref{se:system}. Both exact and asymptotic achievable rate expressions are derived in Section \ref{se:achievable_rate}. Moreover, the power-scaling laws for the considered system is presented in Section \ref{se:achievable_rate}. In Section \ref{se:numerical_results}, a simple power allocation scheme is proposed to compensate for the rate loss. Furthermore, numerical results are provided in Section \ref{se:numerical_results} to illustrate the effect of various system parameters on the achievable rate and energy efficiency. Finally, key findings are concluded in Section \ref{se:conclusion}. Most of the mathematical proofs are given in Appendices A and B.

\subsection{Notations}
In this paper, $\boldsymbol{\rm x}$ and $\boldsymbol{\rm X}$ in bold typeface are used to represent vectors and matrices, respectively, while scalars are presented in normal typeface, such as $x$. We use $\boldsymbol{\rm X}^T$ and $\boldsymbol{\rm X}^H$ to represent the transpose and conjugate transpose of a matrix $\boldsymbol{\rm X}$, respectively. $\boldsymbol{\rm I}_N$ stands for an $N \times N$ identity matrix, and $\|\boldsymbol{\rm X}\|_F$ denotes the Frobenius norm of a matrix $\boldsymbol{\rm X}$. Furthermore, $\mathbb{E}\{\cdot\}$ denotes the expectation operator, and $\boldsymbol{\rm x} \sim \mathcal{CN}(\boldsymbol{\rm m},\sigma^2 \boldsymbol{\rm I} )$ represents a circularly symmetric complex Gaussian stochastic vector with mean vector $\boldsymbol{\rm m}$ and covariance matrix $\sigma^2 \boldsymbol{\rm I}$. Finally, ${{\rm{diag}}\left({\boldsymbol{\rm X}}\right)}$ denotes a diagonal matrix by keeping only the diagonal elements of matrix $\boldsymbol{\rm X}$.

\section{System Model}\label{se:system}
Let us consider a multipair relaying system with $K$ single-antenna user pairs, denoted as $S_k$ and $D_k$,  $k=1, \ldots, K$, applying the relay to exchange information with each other. We assume that the direct links between $S_k$ and $D_k$ do not exist because of large obstacles or severe shadowing. The large-scale relay is equipped with $M$ pairs of antennas, namely receive antennas and transmit antennas, with mixed-resolution ADCs and DACs. In the mixed-ADC/DAC architecture, only $M_0$ pairs of costly high-resolution ADCs and DACs are connected to $M_0$ relay antennas, while the remaining $M_1$ ($=M-M_0$) pairs of less expensive low-resolution ADCs and DACs are connected to $M_1$ relay antennas. Furthermore, we use $\kappa  \buildrel \Delta \over = {{{M_0}} \mathord{\left/{\vphantom {{{M_0}} M}} \right.\kern-\nulldelimiterspace} M}$ $\left( {0 \le \kappa  \le 1} \right)$ to denote the fraction of high-resolution ADCs and DACs in the mixed architecture. The low-resolution ADCs bring about severe quantization errors for data reception and the low-resolution DACs lead to obvious signal distortion for data transmission. Therefore, the correlation of the quantization noise is taken into account for the multipair mMIMO relaying system. Furthermore, we assume that the relay operates in half-duplex mode, so it cannot receive and transmit signals simultaneously. Hence, information transmission from $S_k$ to $D_k$ is completed in two time slots. In the first time slot, the $K$ users in the source set $S_k$ transmit ${\boldsymbol {\rm {x}}_{\rm S}} \in \mathbb{C}^{K \times 1}$ data to the relay independently, and in the next time slot the relay transmits the correlated-quantized signals $\boldsymbol{\tilde {\rm x}}_{\rm R} \in \mathbb{C}^{M \times 1} $ to $K$ users in the destination set $D_k$. The received signal $\boldsymbol{{\rm y}}_{\rm R} \in \mathbb{C}^{M \times 1} $ at the relay and the received signal $\boldsymbol{{\rm y}}_{\rm D} \in \mathbb{C}^{K \times 1} $ can be respectively given by
\begin{align}
\boldsymbol{{\rm y}}_{\rm R} &={{{\boldsymbol{\rm G}_{\rm {SR}}}\boldsymbol{\rm P}_{\rm S}}^{{1 \mathord{\left/
 {\vphantom {1 2}} \right.
 \kern-\nulldelimiterspace} 2}}{\boldsymbol {\rm {x}}_{\rm S}} + {\boldsymbol {\rm {n}}_{\rm {R}}}},\label{eq1}\\
{{\boldsymbol {\rm{y}}}_{D}} &=\gamma {\boldsymbol {\rm{G}}}_{{\rm{RD}}}^T{{\boldsymbol {\rm{\tilde x}}}_{{\rm{R}}}}+ {{\boldsymbol{\rm{n}}}_{D}},\label{eq2}
\end{align}
where $\gamma$ is the normalization factor in order to make the total power at the relay constrained to $p_{\rm R}$, i.e., ${\mathbb{E}}\left\{ {{{\left\| {\gamma {{{\boldsymbol {\rm{\tilde x}}}}_{\rm{R}}}} \right\|}^2}} \right\} = {p_{\rm{R}}}$. Moreover, $\boldsymbol {\rm{P}}_{\rm S}$ is a diagonal matrix representing the transmit power of the $K$ source users and its $k$th element is given by ${\left[ {{{\boldsymbol {\rm{P}}}_{S}}} \right]_{kk}} = {p_{S,k}}$. $ \boldsymbol{\rm n}_{\rm R}\sim \mathcal{CN}\left( {\boldsymbol{0},{{\boldsymbol{\rm I}}_{M}}} \right)$, $ \boldsymbol{\rm n}_{\rm D}\sim \mathcal{CN}\left( {\boldsymbol{0},{{\boldsymbol{\rm I}}_{K}}} \right)$ denote the additive white Gaussian noise (AWGN) matrix with independently and identically distributed (i.i.d.) components following the distribution $ \mathcal{CN}\left( 0,1 \right)$. We further follow the general assumption that the transmit signal vector $\boldsymbol{\rm x_{\rm S}}$  is Gaussian distributed. The matrices ${{\boldsymbol {\rm{G}}}_{{\rm{SR}}}} = \left[ {{\boldsymbol {\rm g}_{{\rm{SR}},1}}, \ldots ,{\boldsymbol{\rm g}_{{\rm{SR}},K}}} \right]$ and ${{\boldsymbol {\rm{G}}}_{{\rm{RD}}}^T} = \left[ {{\boldsymbol {\rm g}_{{\rm{RD}},1}}, \ldots ,{\boldsymbol{\rm g}_{{\rm{RD}},K}}} \right]^T$  refer to the Rayleigh fading channels from the $K$ sources to the relay with ${\boldsymbol{\rm g}_{{\rm{SR}},k}}\sim \mathcal{CN}\left( {\boldsymbol{0},{\beta_{{\rm{SR}},k}}{{\boldsymbol{\rm I}}_{M}}} \right)$ and the channels from the relay to the $K$ destinations with ${\boldsymbol{\rm g}_{{\rm{RD}},k}}\sim \mathcal{CN}\left( {\boldsymbol{0},{\beta_{{\rm{RD}},k}{\boldsymbol{\rm I}}_{M}}} \right)$, respectively. The terms $\beta_{\rm{SR},k}$ and $\beta_{\rm{RD},k}$ stand for the large-scale fading and are assumed to be known at the relay.

Furthermore, we define $\boldsymbol{\rm G}_{\rm {SR0}}$ as the $M_0 \times K$ channel matrix from the $K$ sources to the $M_0$ relay antennas connected with high-resolution ADCs, and $\boldsymbol{\rm G}_{\rm {SR1}}$ as the $M_1 \times K$ channel matrix from the $K$ sources to the remained $M_1$ relay antennas connected with low-resolution ADCs. Therefore, we have
\begin{equation}\label{eq3}
{{\boldsymbol {\rm{G}}}_{{\rm{SR}}}} = \left[ {\begin{array}{*{20}{c}}
{{{\boldsymbol{\rm{G}}}_{{\rm{SR0}}}}}\\
{{{\boldsymbol{\rm{G}}}_{{\rm{SR1}}}}}
\end{array}} \right].
\end{equation}
Similarly, we can also define $\boldsymbol{\rm G}_{\rm {RD0}}^T$ as the $ K \times M_0$ channel matrix from the $M_0$ relay transmit antennas connected with high-resolution DACs to the $K$ destinations, and $\boldsymbol{\rm G}_{\rm {RD1}}^T$ as the $M_1 \times K$ channel matrix from the $M_1$ relay transmit antennas connected with low-resolution DACs to the $K$ destinations. Then, $\boldsymbol{\rm G}_{\rm {RD}}$ can be expressed as
\begin{equation}\label{eq4}
{{\boldsymbol {\rm{G}}}_{{\rm{RD}}}} = \left[ {\begin{array}{*{20}{c}}
{{{\boldsymbol{\rm{G}}}_{{\rm{RD0}}}}}\\
{{{\boldsymbol{\rm{G}}}_{{\rm{RD1}}}}}
\end{array}} \right].
\end{equation}
With the help of \eqref{eq3} and \eqref{eq4}, (\ref{eq1}) and (\ref{eq2}) can be rewritten as
\begin{align}
 \boldsymbol{{\rm y}}_{\rm R} &= \left[ {\begin{array}{*{20}{c}}
\boldsymbol{{\rm y}}_{\rm {R0}} \\
\boldsymbol{{\rm y}}_{\rm {R1}}
\end{array}} \right] = \left[ {\begin{array}{*{20}{c}}
{{{\boldsymbol{\rm G}_{\rm {SR0}}}\boldsymbol{\rm P}_{\rm S}}^{{1 \mathord{\left/
 {\vphantom {1 2}} \right.
 \kern-\nulldelimiterspace} 2}}{\boldsymbol {\rm {x}}_{\rm S}} + {\boldsymbol {\rm {n}}_{\rm {R0}}}}\\
{{\boldsymbol{\rm G}_{\rm {SR1}}}\boldsymbol{\rm P}_{\rm S}}^{{1 \mathord{\left/
 {\vphantom {1 2}} \right.
 \kern-\nulldelimiterspace} 2}}{\boldsymbol {\rm {x}}_{\rm S}} + {\boldsymbol {\rm {n}}_{\rm {R1}}}
\end{array}} \right],\label{eq5}\\
{{\boldsymbol {\rm{y}}}_{D}} &= \left[ {\begin{array}{*{20}{c}}
{{{\boldsymbol {\rm{y}}}_{{\rm{D0}}}}}\\
{{{\boldsymbol {\rm{y}}}_{{\rm{D1}}}}}
\end{array}} \right]   = \gamma {\boldsymbol {\rm{G}}}_{{\rm{RD0}}}^T{{\boldsymbol {\rm{\tilde x}}}_{{\rm{R0}}}} + \gamma {\boldsymbol {\rm{G}}}_{{\rm{RD1}}}^T{{\boldsymbol {\rm{\tilde x}}}_{{\rm{R1}}}} + {{\boldsymbol{\rm{n}}}_{D}},\label{eq6}
\end{align}
where ${{\boldsymbol {\rm{y}}}_{\rm{R0}}}$ denotes the first $M_0$ rows of the overall received signals vector ${\boldsymbol {\rm{y}}}_{\rm{R}}$, and ${{\boldsymbol {\rm{y}}}_{\rm{R1}}}$ denotes the rest $M_1$ rows of ${{\boldsymbol {\rm{y}}}_{\rm{R}}}$. Without loss of generality, the notations ${{\boldsymbol {\rm{y}}}_{\rm{D0}}}$, ${{\boldsymbol {\rm{y}}}_{\rm{D1}}}$, ${\boldsymbol {\rm {n}}_{\rm {R0}}}$, ${\boldsymbol {\rm {n}}_{\rm {R1}}}$, ${{{{\boldsymbol {\rm{\tilde x}}}}_{{\rm{R0}}}}}$ and ${{{{\boldsymbol {\rm{\tilde x}}}}_{{\rm{R1}}}}}$ can also be explained in a similar way.

\subsection{Quantization with Mixed-Resolution ADCs}
For the mixed-ADC architecture, the quantized received signal at the relay can be written as
\begin{equation}\label{eq7}
{\boldsymbol{\tilde{\rm y}}_{\rm R}} = \left[ {\begin{array}{*{20}{c}}
{\boldsymbol{\tilde{\rm y}}_{\rm R0}}\\
{\boldsymbol{\tilde{\rm y}}_{\rm R1}}
\end{array}} \right] = \left[ {\begin{array}{*{20}{c}}
{\boldsymbol{\rm y}_{\rm R0}}\\
{\mathbb{Q}\left( {{\boldsymbol{\rm y}_{\rm R1}}} \right)}
\end{array}} \right],
\end{equation}
where $\mathbb{Q}\left( \cdot \right)$ is the scalar quantization function, ${\boldsymbol{\tilde{\rm y}}_{\rm R0}}$ denotes the quantized received signals at the output of $M_0$ high-resolution ADCs, and ${\boldsymbol{\tilde{\rm y}}_{\rm R1}}$ is the quantized received signals at the output of $M_1$ low-resolution ADCs. According to the AQNM \cite[Eq.  (1)]{orhan2015low}, the quantization operation can be expressed as
\begin{equation}\label{eq8}
{{\boldsymbol{\rm{\tilde y}}}_{{\rm{R1}}}} = \mathbb{Q}\left( {{{\boldsymbol{\rm{y}}}_{{\rm{R1}}}}} \right) = \alpha {{\boldsymbol{\rm{y}}}_{{\rm{R1}}}} + {{\boldsymbol{\rm{n}}}_{{{\rm{q}}_{\rm{a}}}}},
\end{equation}
where ${{\boldsymbol{\rm{n}}}_{{{\rm{q}}_{\rm{a}}}}}$ refers to the additive Gaussian quantization noise vector which is uncorrelated with ${{\boldsymbol{\rm{y}}}_{{\rm{R1}}}}$, {and $\alpha$ denotes a linear gain {given} by \cite[Eq.  (13)]{fletcher2007robust}
\begin{equation}\label{eq9}
\alpha = 1-\rho = 1- {{\mathbb{E}\left\{ {{{\left\| {{{{\boldsymbol{\rm{\tilde y}}}}_{\rm R1}} - {{\boldsymbol{\rm{y}}}_{\rm R1}}} \right\|}^2}} \right\}}}/{{\mathbb{E}\left\{ {{{\left\| {{{{\boldsymbol{\rm{\tilde y}}}}_{\rm R1}}} \right\|}^2}} \right\}}},
\end{equation}
{with $\rho$ as} the distortion factor of the low-resolution ADCs. }The exact values of $\rho$ are given in Table I with respect to different resolution bits \cite{max1960quantizing}. For large quantization bits (e.g., $b>5$), the distortion factor $\rho$ can be approximated as $\rho  \approx \frac{{\pi \sqrt 3 }}{2}{2^{ - 2b}}$ \cite{max1960quantizing}.
With the help of (\ref{eq5}), (\ref{eq8}) and (\ref{eq9}), the covariance matrix of ${{\boldsymbol{\rm{n}}}_{{{\rm{q}}_{\rm{a}}}}}$ is  {expressed as}
\begin{equation}\label{eq11}
\boldsymbol {\rm R}_{{\boldsymbol{\rm{n}}}_{{{\rm{q}}_{\rm{a}}}}}=\alpha \rho {\rm{diag}}\left( {\boldsymbol {\rm G}_{\rm SR1}{{\boldsymbol{\rm{P}}}_{S}}{\boldsymbol {\rm G}_{\rm SR1}^H} + {\boldsymbol {{\rm I}}_{{M_1}}}} \right).
\end{equation}
Moreover, (\ref{eq7}) can be rewritten as
\begin{equation}\label{eq12}
{\boldsymbol{\tilde{\rm y}}_{\rm R}} = \left[ {\begin{array}{*{20}{c}}
{\boldsymbol{\tilde{\rm y}}_{\rm R0}}\\
{\boldsymbol{\tilde{\rm y}}_{\rm R1}}
\end{array}} \right] = \left[ {\begin{array}{*{20}{c}}
{{{\boldsymbol{\rm G}_{\rm {SR0}}}\boldsymbol{\rm P}_{\rm S}}^{{1 \mathord{\left/
 {\vphantom {1 2}} \right.
 \kern-\nulldelimiterspace} 2}}{\boldsymbol {\rm {x}}_{\rm S}} \!+\!{\boldsymbol {\rm {n}}_{\rm {R0}}}}\\
\alpha{{{\boldsymbol{\rm G}_{\rm {SR1}}}\boldsymbol{\rm P}_{\rm S}}^{{1 \mathord{\left/
 {\vphantom {1 2}} \right.
 \kern-\nulldelimiterspace} 2}}{\boldsymbol {\rm {x}}_{\rm S}} \!+\! \alpha{\boldsymbol {\rm {n}}_{\rm {R1}}} \!+\! {{\boldsymbol{\rm{n}}}_{{{\rm{q}}_{\rm{a}}}}}}
\end{array}} \right].
\end{equation}

\begin{table}\normalsize\label{table1}
\caption{\small{Distortion Factors For Different Quantization Bits{.}}}
\centering
\begin{tabular}{cccccc}
\hline
$b$&1&2&3&4&5\\
\hline
$\rho$ &0.3634&0.1175&0.03454&0.009497&0.002499\\
\hline
\end{tabular}
\end{table}

\subsection{Maximum Ratio {(MR)} Processing at the Relay}
We assume that the relay adopts a simple amplify-and-forward (AF)\footnote{{The AF protocol is considered herein due to its lower implementation complexity compared with the decode-and-forward (DF) protocol \cite{zhang2017spectral}.}} protocol to process the quantized received signals, yielding
\begin{equation}\label{eq13}
{\boldsymbol{\rm x}_{\rm R}} = \boldsymbol{\rm W}{\boldsymbol{\tilde{\rm y}}_{\rm R}},
\end{equation}
where $\boldsymbol{\rm W}={\boldsymbol {\rm{G}}}_{{\rm{RD}}}^*{\boldsymbol{\rm{G}}}_{{\rm{SR}}}^H$ denotes the {MR} processing. The MR processing is used at the relay due to its low-complexity, being suitable for the low-cost multipair mMIMO relaying system. {Furthermore, according to some research, the MR processing can achieve similar performance as {zero-forcing receiver/zero-forcing transmission (ZFR/ZFT) or minimum mean square error (MMSE)}.}
By applying (\ref{eq3}) and (\ref{eq4}), (\ref{eq13}) is rewritten as
\begin{equation}\label{eq14}
\begin{aligned}
 \boldsymbol{{\rm x}}_{\rm R} =\left[ {\begin{array}{*{20}{c}}
{{\boldsymbol {\rm G}_{\rm RD0}^*}{\boldsymbol {\rm G}_{\rm SR0}^H}{\boldsymbol{\tilde{\rm y}}_{\rm R0}} \!+\! {\boldsymbol {\rm G}_{\rm RD0}^*}{\boldsymbol {\rm G}_{\rm SR1}^H}{\boldsymbol{\tilde{\rm y}}_{\rm R1}}}\\
{{\boldsymbol {\rm G}_{\rm RD1}^*}{\boldsymbol {\rm G}_{\rm SR0}^H}{\boldsymbol{\tilde{\rm y}}_{\rm R0}} \!+\! {\boldsymbol {\rm G}_{\rm RD1}^*}{\boldsymbol {\rm G}_{\rm SR1}^H}{\boldsymbol{\tilde{\rm y}}_{\rm R1}}}
\end{array}} \right].
\end{aligned}
\end{equation}

\subsection{Quantization with Mixed-Resolution DACs}
{For simplicity, we assume that the DACs and ADCs have the same resolution. The analysis method can be extended to arbitrary resolution {cases}.} With mixed-DAC architecture at the transmitter, the mixed-ADC/DAC signals from the relay's transmit antennas can be expressed as
\begin{equation}\label{eq15}
{\boldsymbol{\tilde{\rm x}}_{\rm R}} = \left[ {\begin{array}{*{20}{c}}
{\boldsymbol{\tilde{\rm x}}_{\rm R0}}\\
{\boldsymbol{\tilde{\rm x}}_{\rm R1}}
\end{array}} \right] = \left[ {\begin{array}{*{20}{c}}
{\boldsymbol{\rm x}_{\rm R0}}\\
{\mathbb{Q}\left( {{\boldsymbol{\rm x}_{\rm R1}}} \right)}
\end{array}} \right]= \left[ {\begin{array}{*{20}{c}}
{\boldsymbol{\rm x}_{\rm R0}}\\
 \alpha {{\boldsymbol{\rm{x}}}_{{\rm{R1}}}} +{{\boldsymbol{\rm{n}}}_{{{\rm{q}}_{D}}}}
\end{array}} \right],
\end{equation}
where $\alpha$ is the distortion factor of the low-resolution DACs, and ${{\boldsymbol{\rm{n}}}_{{{\rm{q}}_{D}}}}$ denotes the quantization noise of low-resolution DACs, which is uncorrelated with ${{\boldsymbol{\rm{x}}}_{{\rm{R1}}}}$. Note that ${\boldsymbol{\tilde{\rm x}}_{\rm R0}} = {\boldsymbol{\rm x}_{\rm R0}}$ because of using high-resolution DACs. Similar as for \eqref{eq11}, we can derive the covariance matrix of ${{\boldsymbol{\rm{n}}}_{{{\rm{q}}_{D}}}}$ as
\begin{equation}\label{eq16}
\boldsymbol {\rm R}_{{\boldsymbol{\rm{n}}}_{{{\rm{q}}_{D}}} }=\alpha \rho {\rm{diag}}\left(\boldsymbol {\rm R}_{{\boldsymbol{\rm{x}}}_{{{\rm{R1}}}} } \right),
\end{equation}
where $\boldsymbol {\rm R}_{{\boldsymbol{\rm{x}}}_{{{\rm{R1}}}} }$ is the covariance matrix of ${{\boldsymbol{\rm{x}}}_{{\rm{R1}}}}$. According to (\ref{eq14}), $\boldsymbol {\rm R}_{{\boldsymbol{\rm{x}}}_{{{\rm{R1}}}} }$ can be written as
\begin{equation}\label{eq17}
\begin{aligned}
{\boldsymbol {\rm R}_{{\boldsymbol{\rm{x}}}_{{{\rm{R1}}}}}}
&= {\boldsymbol{\rm{G}}}_{{\rm{RD1}}}^*{\boldsymbol{\rm{G}}}_{{\rm{SR0}}}^H{{\boldsymbol{\rm{R}}}_{{{{\boldsymbol{\rm{\tilde y}}}}_{{\rm{R0}}}}{{{\boldsymbol{\rm{\tilde y}}}}_{{\rm{R0}}}}}}{{\boldsymbol{\rm{G}}}_{{\rm{SR0}}}}{\boldsymbol{\rm{G}}}_{{\rm{RD1}}}^T \\
&+ {\boldsymbol{\rm{G}}}_{{\rm{RD1}}}^*{\boldsymbol{\rm{G}}}_{{\rm{SR0}}}^H{{\boldsymbol{\rm{R}}}_{{{{\boldsymbol{\rm{\tilde y}}}}_{{\rm{R0}}}}{{{\boldsymbol{\rm{\tilde y}}}}_{{\rm{R1}}}}}}{{\boldsymbol{\rm{G}}}_{{\rm{SR1}}}}{\boldsymbol{\rm{G}}}_{{\rm{RD1}}}^T \\
&+ {\boldsymbol{\rm{G}}}_{{\rm{RD1}}}^*{\boldsymbol{\rm{G}}}_{{\rm{SR1}}}^H{{\boldsymbol{\rm{R}}}_{{{{\boldsymbol{\rm{\tilde y}}}}_{{\rm{R1}}}}{{{\boldsymbol{\rm{\tilde y}}}}_{{\rm{R0}}}}}}{{\boldsymbol{\rm{G}}}_{{\rm{SR0}}}}{\boldsymbol{\rm{G}}}_{{\rm{RD1}}}^T\\
&+ {\boldsymbol{\rm{G}}}_{{\rm{RD1}}}^*{\boldsymbol{\rm{G}}}_{{\rm{SR1}}}^H{{\boldsymbol{\rm{R}}}_{{{{\boldsymbol{\rm{\tilde y}}}}_{{\rm{R1}}}}{{{\boldsymbol{\rm{\tilde y}}}}_{{\rm{R1}}}}}}{{\boldsymbol{\rm{G}}}_{{\rm{SR1}}}}{\boldsymbol{\rm{G}}}_{{\rm{RD1}}}^T,
\end{aligned}
\end{equation}
where
\begin{align}
{{\boldsymbol{\rm{R}}}_{{{{\boldsymbol{\rm{\tilde y}}}}_{{\rm{R0}}}}{{{\boldsymbol{\rm{\tilde y}}}}_{{\rm{R0}}}}}}&={\boldsymbol {\rm G}_{\rm SR0}{{\boldsymbol{\rm{P}}}_{S}}{\boldsymbol {\rm G}_{\rm SR0}^H} + {\boldsymbol {{\rm I}}_{{M_0}}}} ,\label{eq18}\\
{{\boldsymbol{\rm{R}}}_{{{{\boldsymbol{\rm{\tilde y}}}}_{{\rm{R0}}}}{{{\boldsymbol{\rm{\tilde y}}}}_{{\rm{R1}}}}}}&=\alpha{\boldsymbol {\rm G}_{\rm SR0}{{\boldsymbol{\rm{P}}}_{S}}{\boldsymbol {\rm G}_{\rm SR1}^H}} ,\label{eq19}\\
{{\boldsymbol{\rm{R}}}_{{{{\boldsymbol{\rm{\tilde y}}}}_{{\rm{R1}}}}{{{\boldsymbol{\rm{\tilde y}}}}_{{\rm{R0}}}}}}&=\alpha{\boldsymbol {\rm G}_{\rm SR1}{{\boldsymbol{\rm{P}}}_{S}}{\boldsymbol {\rm G}_{\rm SR0}^H}} ,\label{eq20}\\
{{\boldsymbol{\rm{R}}}_{{{{\boldsymbol{\rm{\tilde y}}}}_{{\rm{R1}}}}{{{\boldsymbol{\rm{\tilde y}}}}_{{\rm{R1}}}}}} &= \alpha^2\left({\boldsymbol {\rm G}_{\rm SR1}{{\boldsymbol{\rm{P}}}_{S}}{\boldsymbol {\rm G}_{\rm SR1}^H} + {\boldsymbol {{\rm I}}_{{M_1}}}}\right)\notag \\
&+\alpha \rho {\rm{diag}}\left( {\boldsymbol {\rm G}_{\rm SR1}{{\boldsymbol{\rm{P}}}_{S}}{\boldsymbol {\rm G}_{\rm SR1}^H} + {\boldsymbol {{\rm I}}_{{M_1}}}} \right). \label{eq21}
\end{align}
Consequently, the normalization factor $\gamma$ can be expressed as
\begin{equation}\label{eq22}
\gamma = \sqrt { {{{p_R}}}/{{\mathbb{E}}\left\{ {{{\left\| {{{{\boldsymbol {\rm{\tilde x}}}}_{\rm{R}}}} \right\|}^2}} \right\}}}.
\end{equation}

\newtheorem{lemma}{Lemma}
\begin{lemma} \label{lemma1}
For mixed-ADC/DAC multipair mMIMO relaying systems, the expectation of the total transmit power at the relay can be expressed as
\begin{align}\label{eq23}
{{\mathbb{E}}\left\{ {{{\left\| {{{{\boldsymbol {\rm{\tilde x}}}}_{\rm{R}}}} \right\|}^2}} \right\}}  =\mu \left( {{M_0} + \alpha {M_1}} \right),
\end{align}
where $\mu$ is given by
\begin{equation}\label{eq24}
\begin{aligned}
& \mu   =  \alpha \left( {1  -  \alpha } \right){M_1}\sum\limits_{k = 1}^K {{p_{{S},k}}\beta _{{\rm{SR}},k}^2{\beta _{{\rm{RD}},k}}}.
\end{aligned}
\end{equation}
\end{lemma}

\begin{IEEEproof}
Please refer to Appendix \ref{ap1}.
\end{IEEEproof}

Substituting (\ref{eq23}) into (\ref{eq22}), the normalization factor $\gamma$ can be {obtained} as
\begin{equation}\label{eq25}
\gamma = \sqrt { {{{p_R}}}/{{\mu \left( {{M_0} + \alpha {M_1}} \right)}}}.
\end{equation}
With the normalization factor in hand, we can derive the achievable rate in the following section.

\section{Achievable Rate Analysis}\label{se:achievable_rate}

\subsection{Exact Achievable Rate Analysis}
It is assumed that the destination $D_k$ applies only statistical CSI to decode the signal, due to the reason that instantaneous CSI leads to excessive high computational complexity for large antenna arrays in a practical mMIMO system. Combining (\ref{eq1}), (\ref{eq2}), (\ref{eq12}), (\ref{eq14}) and (\ref{eq15}), the received signal at the destination $D_k$ can be expressed as
\begin{equation}\label{eq26}
{y_{{D},k}}=\underbrace{\sqrt {{p_{{S},k}}}  {\mathbb{E}}\left\{ {{T_{k,k}}} \right\}{x_{{S},k}}}_{\text{desired signal}}
+\underbrace {\tilde n_{{D},k}}_{\text{effective noise}},
\end{equation}
where
\begin{equation}\notag
\begin{aligned}
&{ T_{i,j}}\!=\!\gamma{\boldsymbol{\rm g}_{{\rm{RD0}},i}^T{\boldsymbol{\rm G}}_{{\rm{RD0}}}^*{\boldsymbol{\rm G}}_{{\rm{SR0}}}^H{\boldsymbol{\rm g}_{{\rm{SR0}},j}}}\!+\!\gamma\alpha \boldsymbol{\rm g}_{{\rm{RD0}},i}^T{\boldsymbol{\rm G}}_{{\rm{RD0}}}^*{\boldsymbol{\rm G}}_{{\rm{SR1}}}^H{ \boldsymbol{\rm g}_{{\rm{SR1}},j}} \\
&+ \gamma\alpha  \boldsymbol{\rm g}_{{\rm{RD1}},i}^T{\boldsymbol{\rm G}}_{{\rm{RD1}}}^*{\boldsymbol{\rm G}}_{{\rm{SR0}}}^H{ \boldsymbol{\rm g}_{{\rm{SR0}},j}} \!+\! \gamma{{\alpha ^2}  \boldsymbol{\rm g}_{{\rm{RD1}},i}^T{\boldsymbol{\rm G}}_{{\rm{RD1}}}^*{\boldsymbol{\rm G}}_{{\rm{SR1}}}^H{ \boldsymbol{\rm g}_{{\rm{SR1}},j}}},
\end{aligned}
\end{equation}
and
\begin{align}\label{eq28}
&{{\tilde n}_{D,k}}=\underbrace{\sqrt {{p_{{S},k}}} \left( {{T_{k,k}} - {\mathbb{E}}\left\{ {{T_{k,k}}} \right\}} \right){x_{{S},k}}}_{\text{estimation error}} + \underbrace{\sum\limits_{i \ne k} {{T_{k,i}}{x_{{S},i}}}}_{\text{inter-pair interference}}\notag \\
&+ \underbrace{\gamma\left( { {\boldsymbol{\rm{g}}}_{{\rm{RD0}},k}^T{\boldsymbol {\rm{G}}}_{{\rm{RD0}}}^*{\boldsymbol {\rm{G}}}_{{\rm{SR0}}}^H{{\boldsymbol{\rm{n}}}_{{\rm{R0}}}} +\alpha  {\boldsymbol{\rm{g}}}_{{\rm{RD1}},k}^T{\boldsymbol {\rm{G}}}_{{\rm{RD1}}}^*{\boldsymbol {\rm{G}}}_{{\rm{SR0}}}^H{{\boldsymbol{\rm{n}}}_{{\rm{R0}}}}} \right)}_{\text{noise of high-resolution quantization at the relay}} \notag\\
&\!+\! \underbrace{\gamma\left( {\alpha{\boldsymbol{\rm{g}}}_{{\rm{RD0}},k}^T{\boldsymbol {\rm{G}}}_{{\rm{RD0}}}^*{\boldsymbol {\rm{G}}}_{{\rm{SR1}}}^H{{\boldsymbol{\rm{n}}}_{{\rm{R1}}}} \!+\! {\alpha ^2} {\boldsymbol{\rm{g}}}_{{\rm{RD1}},k}^T{\boldsymbol {\rm{G}}}_{{\rm{RD1}}}^*{\boldsymbol {\rm{G}}}_{{\rm{SR1}}}^H{{\boldsymbol{\rm{n}}}_{{\rm{R1}}}}} \right)}_{\text{noise of low-resolution quantization at the relay}}\notag\\
&+ \underbrace{\gamma\left( {{\boldsymbol{\rm{g}}}_{{\rm{RD0}},k}^T{\boldsymbol {\rm{G}}}_{{\rm{RD0}}}^*{\boldsymbol {\rm{G}}}_{{\rm{SR1}}}^H{{\boldsymbol{\rm{n}}}_{{{\rm{q}}_{\rm{a}}}}} + \alpha{\boldsymbol{\rm{g}}}_{{\rm{RD1}},k}^T{\boldsymbol {\rm{G}}}_{{\rm{RD1}}}^*{\boldsymbol {\rm{G}}}_{{\rm{SR1}}}^H{{\boldsymbol{\rm{n}}}_{{{\rm{q}}_{\rm{a}}}}}} \right)}_{\text{quantization noise of ADCs}} \notag\\
&+ \underbrace{\gamma {\boldsymbol{\rm{g}}}_{{\rm{RD1}},k}^T{{\boldsymbol{\rm{n}}}_{{{\rm{q}}_{D}}}}}_{\text{quantization noise of DACs}} + \underbrace{{{ {n}}_{{D},k}}}_{\text{noise}},
\end{align}

where ${{\boldsymbol{\rm{n}}}_{{D},k}}$ is the $k$th element of the vector ${{\boldsymbol{\rm{n}}}_{{D}}}$. { We can derive the signal-to-interference-plus-noise ratio (SINR) expression by using \cite[Eq. (18)]{liu2017pilot}. }
Since the ``desired signal'' and the ``effective noise'' in (\ref{eq26}) are uncorrelated, the exact achievable rate for the $k$-th destination is given in Theorem \ref{theorem1}.

\newtheorem{theorem}{Theorem}
\begin{theorem} \label{theorem1}
{{For mixed-ADC/DAC multipair mMIMO relaying systems and using the capacity bound in \cite{ngo2018total}, the exact closed-form achievable rate of the $k$-th destination is given as}}
\begin{equation}\label{eq29}
{R_k} \!=\! \frac{{{\tau _c} \!-\! 2{\tau _p}}}{{2{\tau _c}}}{\log _2}\left( {1 \!+\! \frac{{{A_k}}}{{{B_k} \!+\! {C_k} \!+\! {D_k} \!+\! {E_k} \!+\! {F_k} \!+\! {G_k} \!+\! 1}}} \right),
\end{equation}
where $\tau_c$ denotes the length (in symbols) of each coherence interval, $\tau_p$ represents the length of the mutually orthogonal pilot sequences, and
\begin{align}
&{A_k}={p_{{S},k}}{\gamma ^2}{\left( {{M_0} + \alpha {M_1}} \right)^4}\beta _{{\rm{SR}},k}^2\beta _{{\rm{RD}},k}^2, \label{eq37}\\
&{B_k}={p_{{S},k}}{\gamma ^2}\left( {{M_0} + {\alpha ^2}{M_1}} \right){\beta _{{\rm{SR}},k}}{\beta _{{\rm{RD}},k}} \big[2{{\left( {{M_0} \!+\! \alpha {M_1}} \right)}^2}  \notag \\
&\times {\beta _{{\rm{SR}},k}}{\beta _{{\rm{RD}},k}}\!+\! \left( {{M_0} + {\alpha ^2}{M_1}} \right)\sum\limits_{m = 1}^K {{\beta _{{\rm SR},m}}{\beta _{{\rm RD},m}}}  \big], \label{eq38}\\
&{C_k}={\gamma ^2}\left( {{M_0} + {\alpha ^2}{M_1}} \right)\sum\limits_{i \ne k} {{p_{{\rm S},i}}}\notag\\
&\times \Big[ {{{\left( {{M_0} + \alpha {M_1}} \right)}^2}\left({\beta _{{\rm SR},k}}\beta _{{\rm RD},k}^2{\beta _{{\rm SR},i}}+{\beta _{{\rm RD},k}}\beta _{{\rm SR},i}^2{\beta _{{\rm RD},i}}\right) }  \notag\\
& + \left( {{M_0} + {\alpha ^2}{M_1}} \right){\beta _{{\rm RD},k}}{\beta _{{\rm SR},i}}\sum\limits_{m = 1}^K  {{{\beta _{{\rm SR},m}}{\beta _{{\rm RD},m}}} } \Big],\label{eq39}\\
&{D_k}={\gamma ^2}{M_0}{\left( {{M_0} + \alpha {M_1}} \right)^2}{\beta _{{\rm SR},k}}\beta _{{\rm RD},k}^2\notag\\
& + {\gamma ^2}{M_0}\left( {{M_0} + {\alpha ^2}{M_1}} \right){\beta _{{\rm RD},k}}\sum\limits_{m = 1}^K {{\beta _{{\rm SR},m}}{\beta _{{\rm RD},m}}} ,\label{eq40}\\
&{E_k}={\gamma ^2}{\alpha^2}{M_1}{\left( {{M_0} + \alpha {M_1}} \right)^2}{\beta _{{\rm SR},k}}\beta _{{\rm RD},k}^2\notag\\
& + {\gamma ^2}{\alpha^2}{M_1}\left( {{M_0} + {\alpha ^2}{M_1}} \right){\beta _{{\rm RD},k}}\sum\limits_{m = 1}^K {{\beta _{{\rm SR},m}}{\beta _{{\rm RD},m}}} , \label{eq41}\\
&{F_k}=\alpha \rho {\gamma ^2}{M_1}{\beta _{{\rm RD},k}}\left\{ \left( {{M_0} + {\alpha ^2}{M_1}} \right) \sum\limits_{m = 1}^K \Big[  {\beta _{{\rm SR},m}}{\beta _{{\rm RD},m}}\right.\notag\\
& \times \left( {\sum\limits_{i = 1}^K {{p_{{\rm S},i}}{\beta _{{\rm SR},i}}}  + {p_{{\rm S},m}}{\beta _{{\rm SR},m}} + 1} \right)  \Big]+  {{\left( {{M_0} + \alpha {M_1}} \right)}^2}\notag\\
& \times \left.{{\beta _{{\rm SR},k}}{\beta _{{\rm RD},k}}\left( {\sum\limits_{i = 1}^K {{p_{{\rm S},i}}{\beta _{{\rm SR},i}}}  + {p_{{\rm S},k}}{\beta _{{\rm SR},k}} + 1} \right)}  \right\},\label{eq42}\\
&{G_k}=\alpha \rho {\gamma ^2}{M_1}\left( {{M_0} + \alpha {M_1}} \right){\beta _{{\rm RD},k}}\Big\{  {{\beta _{{\rm SR},k}}{\beta _{{\rm RD},k}}}   \notag\\
& \times \left( {\sum\limits_{i = 1}^K {{p_{{\rm S},i}}{\beta _{{\rm SR},i}}}  \!+\! 1\!+\! \left( {{M_0} \!+\! \alpha {M_1}} \right){p_{{\rm S},k}}{\beta _{{\rm SR},k}}} \right) \!+\!   \sum\limits_{m = 1}^K {\beta _{{\rm SR},m}}\notag\\
&\times {\beta _{{\rm RD},m}}\left( {\sum\limits_{i = 1}^K {{p_{{\rm S},i}}{\beta _{{\rm SR},i}}}  + 1 + \left( {{M_0} + \alpha {M_1}} \right){p_{{\rm S},m}}{\beta _{{\rm SR},m}}} \right)    \Big\} \notag\\
& \!+\! {\alpha ^2}{\rho ^2}{\gamma ^2}M_1^2{\beta _{{\rm RD},k}}\left( {\sum\limits_{i = 1}^K {{p_{{\rm S},i}}\beta _{{\rm SR},i}^2{\beta _{{\rm RD},i}}} \!+\! {p_{{\rm S},k}}\beta _{{\rm SR},k}^2{\beta _{{\rm RD},k}}} \right).\label{eq43}
\end{align}
\end{theorem}

\begin{IEEEproof}
Please refer to Appendix \ref{ap2}.
\end{IEEEproof}

\newcounter{mytempeqncnt}
\begin{figure*}[b]
\normalsize
\setcounter{mytempeqncnt}{\value{equation}}
\hrulefill
\vspace*{4pt}
\setcounter{equation}{40}
\begin{equation}\label{eq50}
\begin{aligned}
 {a_{ki}} =  \left\{ {\begin{array}{ll}
\frac{\frac{{{\beta _{{\rm SR},i}}}}{{{\beta _{{\rm SR},k}}}}}{{\left( {{M_0}\! +\! \alpha {M_1}} \right)}}\left( {1 \! +\! \frac{{\alpha \rho {M_1}}}{{{{\left( {{M_0} \! +\! \alpha {M_1}} \right)}^2}}} \! +\! \frac{{{\beta _{{\rm SR},i}}{\beta _{{\rm RD},i}}}}{{{\beta _{{\rm SR},k}}{\beta _{{\rm RD},k}}}}\left[ {1 \! +\! \frac{{\alpha \rho {M_1}}}{{{{\left( {{M_0} \! +\! \alpha {M_1}} \right)}^2}}}} \right] \! +\! \frac{1}{{\left( {{M_0} \! +\! \alpha {M_1}} \right)}}\frac{{\sum\limits_{m = 1}^K {{\beta _{{\rm SR},m}}{\beta _{{\rm RD},m}}} }}{{{\beta _{{\rm SR},k}}{\beta _{{\rm RD},k}}}}} \right),& {i \ne k}\\
\frac{1}{{\left( {{M_0} \! +\! \alpha {M_1}} \right)}}\left( {2 \! +\! \frac{{\alpha \rho {M_1}}}{{\left( {{M_0} + \alpha {M_1}} \right)}}\left[ {2 \! +\! \frac{2}{{\left( {{M_0} \! +\! \alpha {M_1}} \right)}} \! +\! \frac{{\alpha \rho {M_1}}}{{{{\left( {{M_0} \! +\! \alpha {M_1}} \right)}^2}}}} \right] + \frac{1}{{\left( {{M_0} \! +\! \alpha {M_1}} \right)}}\frac{{\sum\limits_{m = 1}^K {{\beta _{{\rm SR},m}}{\beta _{{\rm RD},m}}} }}{{{\beta _{{\rm SR},k}}{\beta _{{\rm RD},k}}}}} \right),& {i = k}
\end{array}} \right.
\end{aligned}
\end{equation}
\setcounter{equation}{\value{mytempeqncnt}}
\end{figure*}

\newcounter{mytempeqncnt1}
\begin{figure*}[b]
\normalsize
\setcounter{mytempeqncnt1}{\value{equation}}
\hrulefill
\vspace*{4pt}
\setcounter{equation}{49}
\begin{equation}\label{eq63}
{\hat{R}_k},{R_k} \to \frac{{{\tau _c} - 2{\tau _p}}}{{2{\tau _c}}}{\log _2}\Bigg[ {1 + \frac{{{E_{\rm S}}{E_{\rm R}}{{\left( {\alpha  + \rho \kappa } \right)}^2}}}{{{E_{\rm S}}\left( {\alpha  + \rho \kappa } \right)\sum\limits_{i = 1}^K {\frac{{\beta _{{\rm SR},i}^2{\beta _{{\rm RD},i}}}}{{\beta _{{\rm SR},k}^2\beta _{{\rm RD},k}^2}}}  + \frac{{\sum\limits_{m = 1}^K {{\beta _{{\rm SR},m}}{\beta _{{\rm RD},m}}} }}{{\beta _{{\rm SR},k}^2\beta _{{\rm RD},k}^2}} + \frac{{{E_{\rm R}}\left( {\alpha  + \rho \kappa } \right)}}{{{\beta _{{\rm SR},k}}}}}}} \Bigg].
\end{equation}
\setcounter{equation}{\value{mytempeqncnt1}}
\end{figure*}

With the help of (\ref{eq25}) and after some simplifications, we can derive the compact expression for the sum achievable rate as
\begin{equation}\label{eq44}
{R} = \frac{{{\tau _c} - 2{\tau _p}}}{{2{\tau _c}}}\sum\limits_{k = 1}^K {{{\log }_2}\left( {1 + {\nu _k}} \right)},
\end{equation}
where
\begin{align}
{\nu _k}&= {{{p_{{\rm S},k}}}}/{{{\xi _k}}},\label{eq45}\\
{\xi _k}&=\sum\limits_{i = 1}^K {{p_{{\rm S},i}}{a_{ki}}}  + p_{\rm R}^{ - 1}\left( {\sum\limits_{i = 1}^K {{p_{{\rm S},i}}{b_{ki}}}  + {c_k}} \right) + {d_k},\label{eq46}\\
{b_{ki}} &=\frac{1}{{\left( {{M_0} + \alpha {M_1}} \right)}}\frac{{\beta _{{\rm SR},i}^2{\beta _{{\rm RD},i}}}}{{\beta _{{\rm SR},k}^2\beta _{{\rm RD},k}^2}}\left[ {1 + \frac{{\alpha \rho {M_1}}}{{{{\left( {{M_0} + \alpha {M_1}} \right)}^2}}}} \right]\notag \\
&+ \frac{1}{{{{\left( {{M_0} + \alpha {M_1}} \right)}^2}}}\frac{{{\beta _{{\rm SR},i}}\sum\limits_{m = 1}^K {{\beta _{{\rm SR},m}}{\beta _{{\rm RD},m}}} }}{{\beta _{{\rm SR},k}^2\beta _{{\rm RD},k}^2}},\label{eq47}\\
{c_k} &= \frac{1}{{{{\left( {{M_0} + \alpha {M_1}} \right)}^2}}}\frac{{\sum\limits_{m = 1}^K {{\beta _{{\rm SR},m}}{\beta _{{\rm RD},m}}} }}{{\beta _{{\rm SR},k}^2\beta _{{\rm RD},k}^2}},\label{eq48}\\
{d_k} &= \frac{1}{{\left( {{M_0} + \alpha {M_1}} \right)}}\frac{1}{{{\beta _{{\rm SR},k}}}}+\frac{{\alpha \rho {M_1}}}{{{{\left( {{M_0} + \alpha {M_1}} \right)}^3}}}\frac{1}{{{\beta _{{\rm SR},k}}}} \notag \\
&+ \frac{1}{{{{\left( {{M_0} + \alpha {M_1}} \right)}^2}}}\frac{{\sum\limits_{m = 1}^K {{\beta _{{\rm SR},m}}{\beta _{{\rm RD},m}}} }}{{\beta _{{\rm SR},k}^2{\beta _{{\rm RD},k}}}},\label{eq49}
\end{align}
with $a_{ki}$ given by \eqref{eq50} at the bottom of next page.

\subsection{Asymptotic Analysis}
Note that ${F_k}$ and ${G_k}$ are respectively derived by using exact values of the covariance matrices $\boldsymbol {\rm R}_{{\boldsymbol{\rm{n}}}_{{{\rm{q}}_{\rm{a}}}} }$ and $\boldsymbol {\rm R}_{{\boldsymbol{\rm{n}}}_{{{\rm{q}}_{D}}} }$, which make results in (\ref{eq42}) and (\ref{eq43}) cumbersome. In order to provide more insights into the effect of various parameters on the achievable rate, we consider a large number of antennas and use the law of large numbers. The covariance matrix $\boldsymbol {\rm R}_{{\boldsymbol{\rm{n}}}_{{{\rm{q}}_{\rm{a}}}} }$ in (\ref{eq11}) is given by
\setcounter{equation}{41}
\begin{align}\label{eq51}
\boldsymbol {\rm R}_{{\boldsymbol{\rm{n}}}_{{{\rm{q}}_{\rm{a}}}}} \approx \alpha \rho \left( {\sum\limits_{k = 1}^K {{p_{{\rm S},k}}{\beta _{{\rm SR},k}}}  + 1} \right){\boldsymbol{\rm I}_{{ M_1}}}.
\end{align}
Similarly, the covariance matrix $\boldsymbol {\rm R}_{{\boldsymbol{\rm{n}}}_{{{\rm{q}}_{D}}} }$ is approximated as
\begin{equation}\label{eq52}
\begin{aligned}
\boldsymbol {\rm R}_{{\boldsymbol{\rm{n}}}_{{{\rm{q}}_{D}}}} &\approx {\mathbb{E}}\left\{ \boldsymbol {\rm R}_{{\boldsymbol{\rm{n}}}_{{{\rm{q}}_{D}}}} \right\} =\alpha \rho \mu{\boldsymbol{\rm I}_{{ M_1}}},
\end{aligned}
\end{equation}
where $\mu$ is given by (\ref{eq24}). Note that the expression of the approximate $\boldsymbol {\rm R}_{{\boldsymbol{\rm{n}}}_{{{\rm{q}}_{D}}}}$ is obtained in \eqref{eq101}, in Appendix \ref{ap1}.
Based on the aforementioned discussion, we can derive a more concise close-form approximation for the achievable rate by simplifying ${F_k}$ and ${G_k}$ as in (\ref{eq51}) and (\ref{eq52}), respectively.
{Similar to Theorem 1, we can derive the approximate achievable rate $\hat{R}_k$ by calculating the power expectations of the signal and interference as shown in the following theorem.}

\begin{theorem}\label{theorem2}
For mixed-ADC/DAC multipair mMIMO relaying systems, the approximate achievable rate of the $k$-th destination is
\begin{equation}\label{eq53}
{\hat{R}_k} \!=\! \frac{{{\tau _c} \!-\! 2{\tau _p}}}{{2{\tau _c}}}{\log _2}\left( {1 \!+\! \frac{{{A_k}}}{{{B_k} \!+\! {C_k}\!+\! {D_k} \!+\! {E_k} \!+\! {\hat{F}_k} \!+\! {\hat{G}_k} \!+\! 1}}} \right),
\end{equation}
where ${\hat{G}_k}={\gamma ^2}\alpha\rho\mu{M_1}{\beta _{{\rm RD},k}}$,
\begin{align}\label{eq56}
&{\hat{F}_k}={\gamma ^2}\alpha\rho \left( {\sum\limits_{k = 1}^K {{p_{{\rm S},k}}{\beta _{{\rm SR},k}}}  + 1} \right){M_1}{\beta _{{\rm RD},k}}\Big({\left({ {M_0} + \alpha {M_1}} \right)^2}\notag \\
&\times {\beta _{{\rm SR},k}}\beta _{{\rm RD},k} + \left( {{M_0} + {\alpha ^2}{M_1}} \right)\sum\limits_{m = 1}^K {{\beta _{{\rm SR},m}}{\beta _{{\rm RD},m}}}\Big),
\end{align}
and $A_k$, $B_k$, $C_k$, $D_k$ and $E_k$ are given by (\ref{eq37}), (\ref{eq38}), (\ref{eq39}), (\ref{eq40}), and (\ref{eq41}), respectively.
\end{theorem}

\begin{IEEEproof}
From \eqref{eq119} in Appendix B, we have
\begin{align}\label{eq:f_k}
{\hat{F}_k} =\frac{{{\rho}}}{{{\alpha}}} \left( {\sum\limits_{k = 1}^K {{p_{{\rm S},k}}{\beta _{{\rm SR},k}}}  + 1} \right){E_k}.
\end{align}
Substituting (\ref{eq51}) into \eqref{eq:f_k}, we {obtain} \eqref{eq56}. Similarly, we derive ${\hat{G}_k}={\gamma ^2}\alpha\rho\mu{\mathbb{E}}\left\{ {\left| {{\boldsymbol{\rm g}}_{\rm {RD1},k}^T{\boldsymbol{\rm g}}_{\rm{RD1},k}^*} \right|} \right\}.$
Using the fact that ${\mathbb{ E}}\left\{ {\left| {{\boldsymbol{\rm g}}_{\rm {RD1},k}^T{\boldsymbol{\rm g}}_{\rm{RD1},k}^*} \right|} \right\}={M_1}{\beta _{{\rm SR},k}}$, we can then derive ${\hat{G}_k}$.
\end{IEEEproof}

Following a similar reasoning as in the exact achievable rate analysis, we substitute (\ref{eq24}) and (\ref{eq25}) into Theorem \ref{theorem2} to deduce the compact expression for the approximate achievable rate as
\begin{equation}\label{eq58}
{\hat{R}_k} =  \left({\left({{\tau _c} - 2{\tau _p}}\right)}/{{2{\tau _c}}}\right)\sum\limits_{k = 1}^K {{{\log }_2}\left( {1 + {\hat{\nu} _k}} \right)},
\end{equation}
where
\begin{align}
{\hat{\nu} _k}=& {{{p_{{\rm S},k}}}}/{{{\hat{\xi} _k}}},\label{eq59}\\
{\hat{\xi} _k}=&\sum\limits_{i = 1}^K {{p_{{\rm S},i}}{\hat{a}_{ki}}}  +  \left( {\sum\limits_{i = 1}^K {{p_{{\rm S},i}}{b_{ki}}}  + {c_k}} \right)/p_{\rm R} + {\hat{d}_k},\label{eq60}\\
{\hat{d}_k} = &\frac{1}{{\left( {{M_0} + \alpha {M_1}} \right)} {{\beta _{{\rm SR},k}}}}+ \frac{1}{{{{\left( {{M_0} + \alpha {M_1}} \right)}^2}}}\frac{{\sum\limits_{m = 1}^K {{\beta _{{\rm SR},m}}{\beta _{{\rm RD},m}}} }}{{\beta _{{\rm SR},k}^2{\beta _{{\rm RD},k}}}},\notag
\end{align}
with $\hat{a}_{ki}$ given as
\begin{align}
&{\hat{a}_{ki}}=\frac{1}{{\left( {{M_0} \!+\! \alpha {M_1}} \right)}}\frac{{{\beta _{{\rm SR},i}}}}{{{\beta _{{\rm SR},k}}}}\Big\{  1 \!+\! \frac{{{\beta _{{\rm SR},i}}{\beta _{{\rm RD},i}}}}{{{\beta _{{\rm SR},k}}{\beta _{{\rm RD},k}}}}\notag \\
& \times \left[ {1 \!+\! \frac{{{\alpha ^2}{\rho ^2}M_1^2}}{{{{\left( {{M_0} \!+\! \alpha {M_1}} \right)}^3}}}} \right]+  \frac{1}{{\left( {{M_0} \!+\! \alpha {M_1}} \right)}}\frac{{\sum\limits_{m = 1}^K {{\beta _{{\rm SR},m}}{\beta _{{\rm RD},m}}} }}{{{\beta _{{\rm SR},k}}{\beta _{{\rm RD},k}}}}  \Big\}.\notag
\end{align}
Note that $b_{ki}$ and $c_k$ have been defined in (\ref{eq47}) and (\ref{eq48}), respectively.

It is clear to see from \eqref{eq58} that the approximate achievable rate ${\hat{R}_k}$ increases with the total power of the relay $p_{R}$, and decreases with the transmit power of other sources. The mixed-ADC/DAC multipair mMIMO relaying system is interference-limited, which is consistent with \cite{kong2017multipair}. Moreover, we find that including more low-resolution ADCs and DACs decreases ${\hat{R}_k}$. This is reasonable since the quantization noise increases.

\subsection{Power Scaling Law}
In this subsection, we investigate the potential for power saving in the data transmission phase due to the deployment of a very large antenna array at the relay. Here, let $p_{\rm S}={{{E_{\rm S}}} \mathord{\left/{\vphantom {{{E_{\rm S}}} M}} \right. \kern-\nulldelimiterspace} M}$ (i.e., the power of all sources is the same, $p_{\rm S} =p_{{\rm S},k}$, $k=1,\ldots,K$) and $p_{\rm R}={{{E_{\rm R}}} \mathord{\left/{\vphantom {{{E_{\rm R}}} M}} \right. \kern-\nulldelimiterspace} M}$, where {{the transmit power of the source $E_{S}$ and of the relay $E_{ R}$ are fixed.}} As $M \to \infty $, the exact and approximate achievable rate for the considered system is provided in the following Corollary.

\newtheorem{corollary}{Corollary}
\begin{corollary}  \label{corollary1}
With $p_{\rm S}={{{E_{\rm S}}} \mathord{\left/{\vphantom {{{E_{\rm S}}} M}} \right. \kern-\nulldelimiterspace} M}$, $p_{\rm R}={{{E_{\rm R}}} \mathord{\left/{\vphantom {{{E_{\rm R}}} M}} \right. \kern-\nulldelimiterspace} M}$ and $E_{\rm S}$, $E_{\rm R}$ fixed, the achievable rate limit of mixed-ADC/DAC multipair mMIMO relaying systems is given by \eqref{eq63} at the bottom of this page.
\end{corollary}

\begin{IEEEproof}
We start with the approximate achievable rate ${\hat{R}_k}$. Let $p_{\rm S}={{{E_{\rm S}}} \mathord{\left/{\vphantom {{{E_{\rm S}}} M}} \right. \kern-\nulldelimiterspace} M}$, $p_{\rm R}={{{E_{\rm R}}} \mathord{\left/{\vphantom {{{E_{\rm R}}} M}} \right. \kern-\nulldelimiterspace} M}$, and with the help of (\ref{eq59}) and (\ref{eq60}), we can obtain
\setcounter{equation}{50}
\begin{equation}\label{eq64}
{\hat{\nu} _k}=\frac{{{E_{{\rm S}}}}}{{{E_{{\rm S}}}\sum\limits_{i = 1}^K {{\hat{a}_{ki}}}  + E_{\rm R}^{ - 1}\left( {\sum\limits_{i = 1}^K {{E_{{\rm S}}}{M}{b_{ki}}}  + {M^2}{c_k}} \right) + M{\hat{d}_k}}}.
\end{equation}
As $M \to\infty$, the terms related to $M$ in (\ref{eq64}) are derived as
\begin{align}
&\hat{a}_{ki}\to 0,\label{eq65}\\
&M{b_{ki}}\to {{\beta _{{\rm SR},i}^2{\beta _{{\rm RD},i}}}}/{{\left( {\alpha+ \rho\kappa} \right)}} {{\beta _{{\rm SR},k}^2\beta _{{\rm RD},k}^2}},\label{eq66}\\
&M^2{c_k}\to\frac{1}{{{{\left( {{\alpha+ \rho\kappa}} \right)}^2}}}\frac{{\sum\limits_{m = 1}^K {{\beta _{{\rm SR},m}}{\beta _{{\rm RD},m}}} }}{{\beta _{{\rm SR},k}^2\beta _{{\rm RD},k}^2}},\label{eq67}\\
&M{\hat{d}_k}\to\frac{1}{{\left( {\alpha+\rho\kappa} \right)}}\frac{1}{{{\beta _{{\rm SR},k}}}}\label{eq68}.
\end{align}
Substituting (\ref{eq65}), \eqref{eq66}, \eqref{eq67} and (\ref{eq68}) into (\ref{eq64}), we can derive the limit of $\hat{R}_k$ after some simple mathematical manipulations. Following a similar way, the limit of exact achievable rate $R_k$ can be derived. With $p_{\rm S}={{{E_{\rm S}}} \mathord{\left/{\vphantom {{{E_{\rm S}}} M}} \right. \kern-\nulldelimiterspace} M}$ and $p_{\rm R}={{{E_{\rm R}}} \mathord{\left/{\vphantom {{{E_{\rm R}}} M}} \right. \kern-\nulldelimiterspace} M}$, (\ref{eq45}) can be rewritten as
\begin{equation}\label{eq69}
{\nu_k}=\frac{{{E_{{\rm S}}}}}{{{E_{{\rm S}}}\sum\limits_{i = 1}^K {{a_{ki}}}  \!+\! E_{\rm R}^{ \!-\! 1}\left( {\sum\limits_{i = 1}^K {{E_{{\rm S}}}{M}{b_{ki}}}  \!+\! {M^2}{c_k}} \right) \!+\! M{d_k}}}.
\end{equation}
As $M \to\infty$, the terms related to $M$ in (\ref{eq69}) are given by
\begin{align}
&a_{ki}\to 0,\label{eq70}\\
&M{d_k}\to\frac{1}{{\left( {\alpha+\rho\kappa} \right)}}\frac{1}{{{\beta _{{\rm SR},k}}}}\label{eq710},
\end{align}
and the limits of $M{b_{ki}}$ and $M^2{c_k}$ are given by (\ref{eq66}) and (\ref{eq67}), respectively. Since the limit of $a_{ki}$ and $d_k$ are separately the same as $\hat{a}_{ki}$ and $\hat{d}_k$ with $M\to\infty$, $R_k$ {approaches} the same constant limit as $\hat{R}_k$. After some simplifications, the proof is concluded by deriving (\ref{eq63}).
\end{IEEEproof}

It is clear from (\ref{eq63}) that both exact and approximate results tend to a same constant value with $M \to \infty$. We can find that the proportion of the high-resolution ADCs/DACs $\kappa$ and the distortion factor of the low-resolution ADCs/DACs $\rho$ have effects on the limit rate when scaling down the transmit power proportion to ${{{1}} \mathord{\left/{\vphantom {{{1}} M}} \right.\kern-\nulldelimiterspace} M}$. More specifically, the limit can be improved by increasing $\kappa$. Adopting the fact that $\rho\kappa+\alpha=\left(1-\kappa\right)\alpha+\kappa$ is a monotonic increasing function of $\alpha$, the limit of  achievable rate monotonically increases with $\alpha$, which means that we can boost the achievable rate by using higher quantization bits in the $M_1$ low-resolution ADCs/DACs.

\newtheorem{proposition}{Proposition}
\begin{proposition}  \label{proposition1}
With $p_{\rm S}={{{E_{\rm S}}} \mathord{\left/{\vphantom {{{E_{\rm S}}} M}} \right. \kern-\nulldelimiterspace} M}$, $p_{\rm R}={{{E_{\rm R}}} \mathord{\left/{\vphantom {{{E_{\rm R}}} M}} \right. \kern-\nulldelimiterspace} M}$ and $E_{\rm S} \to 0$, $E_{\rm R}$ fixed, we can derive the factor of the sum rate gap between the mixed-ADC/DAC relay system and the unquantized one as
\begin{equation}
\frac{R_{k}}{R^p_k} \to \frac{{{{\left( {\alpha  + \rho \kappa } \right)}^2}\left( {{E_{\rm{R}}}{\beta _{{\rm SR},k}}\beta _{{\rm RD},k}^2 + \sum\limits_{m = 1}^K {{\beta _{{\rm SR},m}}{\beta _{{\rm RD},m}}} } \right)}}{{\left( {\alpha  + \rho \kappa } \right){E_{\rm{R}}}{\beta _{{\rm SR},k}}\beta _{{\rm RD},k}^2 + \sum\limits_{m = 1}^K {{\beta _{{\rm SR},m}}{\beta _{{\rm RD},m}}} }}.\notag
\end{equation}
\end{proposition}

\begin{proposition} \label{proposition2}
With $p_{\rm S}={{{E_{\rm S}}} \mathord{\left/{\vphantom {{{E_{\rm S}}} M}} \right. \kern-\nulldelimiterspace} M}$, $p_{\rm R}={{{E_{\rm R}}} \mathord{\left/{\vphantom {{{E_{\rm R}}} M}} \right. \kern-\nulldelimiterspace} M}$ and $E_{\rm S}$ fixed, $E_{\rm R} \to 0$, we can derive the factor of the sum rate gap between the mixed-ADC/DAC relay system and the unquantized one as
\begin{equation}\label{addeq2}
\frac{R_{k}}{R^p_k} \to \frac{{{{\left( {\alpha  + \rho \kappa } \right)}^2}\left( {{E_{S}}\sum\limits_{m = 1}^K {\beta _{{\rm SR},m}^2{\beta _{{\rm RD},m}}}  + \sum\limits_{m = 1}^K {{\beta _{{\rm SR},m}}{\beta _{{\rm RD},m}}} } \right)}}{{\left( {\alpha  + \rho \kappa } \right){E_{S}}\sum\limits_{m = 1}^K {\beta _{{\rm SR},m}^2{\beta _{{\rm RD},m}}}  + \sum\limits_{m = 1}^K {{\beta _{{\rm SR},m}}{\beta _{{\rm RD},m}}} }}.
\end{equation}
\end{proposition}

From Propositions \ref{proposition1} and \ref{proposition2}, it is clear that the sum rate gap between the mixed-ADC/DAC relay system and the unquantized system is a constant factor in the low power regime. The factors in the case $E_{\rm S}\to0$ or $E_{\rm R}\to0$ are both related to $\alpha+\rho\kappa$. Using the fact that $\rho\kappa+\alpha=\left(1-\kappa\right)\alpha+\kappa$ is a monotonic increasing function of $\alpha$, we can find that the factors also increase with $\alpha$. For the special case of $\alpha=1$, i.e., $\alpha+\rho\kappa=1$, the achievable rate of the mixed-ADC/DAC system is the same as that of the ideal unquantized system.

\section{Power Allocation}\label{se:power_allocation}
In this section, we try to maximize the sum achievable rate of the mixed-ADC/DAC multipair mMIMO relaying system constrained to a given total sum power $P_{\rm T}$, i.e., $\sum\limits_{k = 1}^K {{p_{{\rm S},k}}}+p_{\rm R} \le P_{\rm T}$, and formulate it as a power allocation problem.

Let us define ${\boldsymbol{\rm p}_{\rm S}}={\left[ {{{p_{{S},1}},\ldots,{p_{{S},K}}}} \right]}^{T}$, the power allocation problem can be expressed as
\begin{align}
\mathcal{P}_1:\mathop {{\rm{maximize}}}\limits_{{\boldsymbol{\rm p}_{\rm S}},p_{\rm R}}\quad&\frac{{{\tau _c} - 2{\tau _p}}}{{2{\tau _c}}}\sum\limits_{k = 1}^K {{{\log }_2}\left( {1 + {\nu _k}} \right)}\label{eq71}\\
\rm{subject}\ \rm{to}\quad&\sum\limits_{k = 1}^K {{p_{{\rm S},k}}}+p_{\rm R} \le P_{\rm T}\label{eq72}\\
&{\boldsymbol{\rm p}_{\rm S}} \ge \boldsymbol{\rm 0},p_{\rm R}\ge 0.\label{eq73}
\end{align}

Since $\log\left({\cdot }\right)$ is a monotonic increasing function, {the} problem $\mathcal{P}_1$ can be reformulated as
\begin{align}
\mathcal{P}_2:\mathop {{\rm{minimize}}}\limits_{{\boldsymbol{\rm p}_{\rm S}},p_{\rm R}}\quad&\prod\limits_{k=1}^K{\left( {1 + {\nu _k}} \right)}^{-1}\label{eq74}\\
\rm{subject}\ \rm{to}\quad&\sum\limits_{k = 1}^K {{p_{{\rm S},k}}}+p_{\rm R} \le P_{\rm T}\label{eq75}\\
&{\boldsymbol{\rm p}_{\rm S}} \ge \boldsymbol{\rm 0},p_{\rm R}\ge 0.\label{eq76}
\end{align}
Problem $\mathcal{P}_2$ is a general nonconvex complementary geometric program (CGP), which can be approximated by solving a sequence of GP problems. After that, we can use standard convex optimization tools (e.g., CVX) to solve the GP problems \cite{boyd2004convex}. {{The detailed steps of the power allocation algorithm are provided in Algorithm \ref{alg1}. Following the successive approximation algorithm in \cite{kong2017multipair}, it is efficient to solve the power allocation problem.}}

\begin{algorithm}[htb]\caption{Successive approximation algorithm for $\mathcal{P}_2$ }\label{alg1}1) \emph{Initialization.} Define a tolerance $\epsilon$ and parameter $\theta$. Set $j=1$ and set the initial value $\tilde{\nu} _k$ according to the SINR expression in Theorem 1 with $p_{S,k}=\frac{P_T}{2K}$ and $p_R=\frac{P_T}{2}$.
2) \emph{iteration $j$.} Compute $\delta_k=\frac{\tilde{\nu} _k}{1+\tilde{\nu} _k}$. Then solve the GP problem $\mathcal{P}_3$:\begin{align}\mathcal{P}_3:\mathop {{\rm{minimize}}}\limits_{{\boldsymbol{\rm p}_{\rm S}},p_{\rm R}}\quad&\prod\limits_{k=1}^K{\nu _k}^{-\delta_k}\label{eqalg1}\\\rm{subject}\ \rm{to}\quad&{\theta^{-1}\tilde{\nu} _k \le \nu_k \le \theta\tilde{\nu} _k,\ k=1,\cdots,K}\label{eqalg2}\\&{\nu _k{p_{S,k}}^{-1}\xi_k \le 1,\ k=1,\cdots,K}\label{eqalg3}\\&{\sum\limits_{k = 1}^K {{p_{{\rm S},k}}}+p_{\rm R} \le P_{\rm T}}\label{eqalg4}\\&{\boldsymbol{\rm p}_{\rm S}} \ge \boldsymbol{\rm 0},p_{\rm R}\ge 0.\label{eqalg5}\end{align}Denote the optimal solutions by ${\nu_k}^{\left(j\right)}$, for $k=1,\cdots,K$.
3) \emph{Stopping criterion}. If ${\rm{max}}_k \left|{{\nu_k}^{\left(j\right)}-{\tilde{\nu}_k}}\right|<\epsilon$, stop; otherwise, go to step 4).
4) \emph{Update initial values}. Set $\nu_k={\nu_k}^{\left(j\right)}$, and $j=j+1$. Go to step 2)\end{algorithm}

%
%
%

\begin{figure}[!t]
\centering
\includegraphics[scale=0.65]{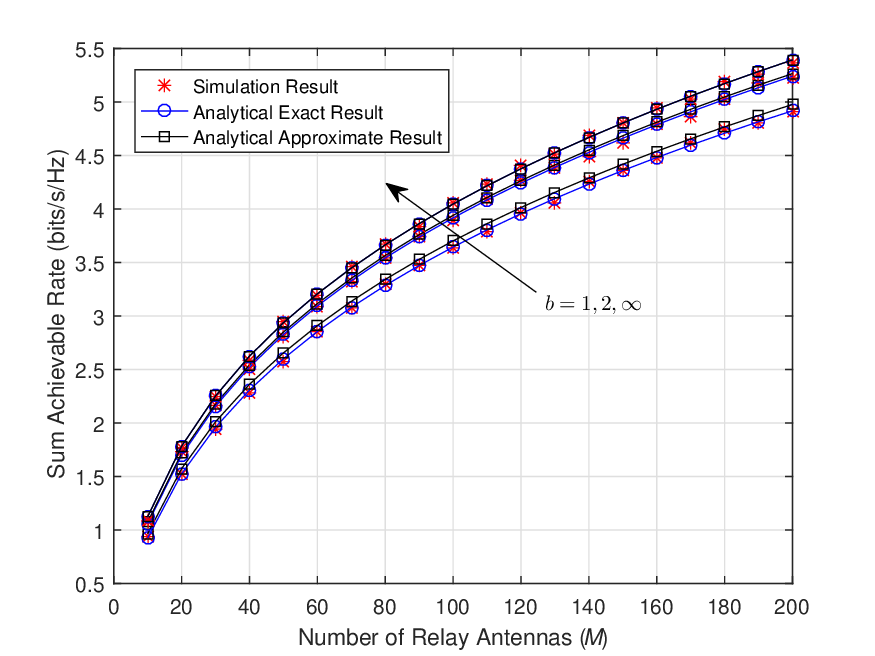}
\caption{Sum achievable rate against the number of relay antennas $M$ for $p_{\rm S}=10$ dB, $p_{\rm R}=10$ dB, $K=10$ and $\kappa=1/2$.}
\label{fig1}
\end{figure}

\section{Numerical Results}\label{se:numerical_results}
In this section, we conduct numerous simulations to verify the accuracy of the analytical results. Apart from that, insights are also provided. Then, the benefit of the proposed power allocation algorithm is demonstrated. Moreover, we investigate the energy efficiency to show the advantage of the mixed-ADC/DAC architecture.

In the Monte Carlo simulation, we assume that the users are distributed in a hexagonal cell with a radius of 1000 meters, while the minimum distance between the users and relay is $r_{\min}=100$ meters.
{{The length of the coherence interval and pilot sequence are set as $\tau_c=20K$ and $\tau_p=K$, respectively.}}
Furthermore, the large-scale fading coefficients are arbitrarily generated by  ${\beta _{{\rm SR},k}} = {z_k}{\left( {{{{r_{{\rm SR},k}}} \mathord{\left/ {\vphantom {{{r_{{\rm SR},k}}} {{r_{\min }}}}} \right. \kern-\nulldelimiterspace} {{r_{\min }}}}} \right)^{ - \alpha}}$ and  ${\beta _{{\rm RD},k}} = {z_k}{\left( {{{{r_{{\rm RD},k}}} \mathord{\left/ {\vphantom {{{r_{{\rm RD},k}}} {{r_{\min }}}}} \right. \kern-\nulldelimiterspace} {{r_{\min }}}}} \right)^{ - \alpha}}$, where $z_k$ is a log-normal random variable with standard derivation 8 dB, ${r_{{\rm SR},k}}$ and ${r_{{\rm RD},k}}$ represent the distances from the sources to the relay and destinations to the relay, respectively, and $\alpha=3.8$ denotes the pathloss exponent.

\subsection{Achievable Rate}
In Fig. \ref{fig1}, the simulated achievable rate, the analytical exact result (\ref{eq44}) and the approximate result (\ref{eq58}) are plotted against the number of relay antennas. It can be seen that the analytical exact and approximate results, as well as simulation results are close to each other, which validates the correctness of our derived expressions. For a small number of antennas at the relay, the sum achievable rates for the cases of $b=1,2,\infty$ matches well with each other. While as the quantization bits $b$ increase, the gap between the approximate and simulated curves becomes small. Finally, better rate performance is achieved with a larger number of quantization bits ($b>1$).

\begin{figure}[!t]
\centering
\includegraphics[scale=0.65]{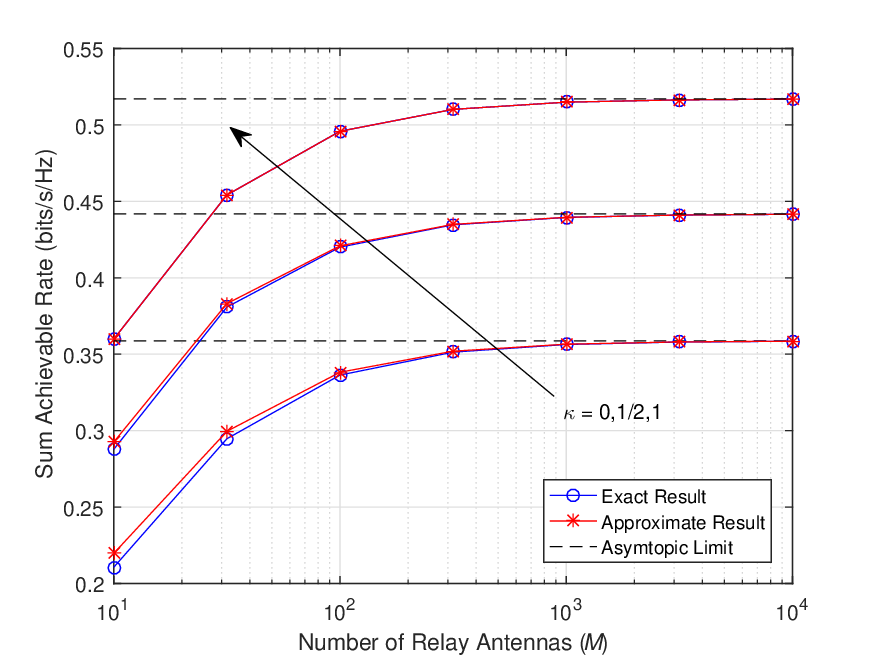}
\caption{Sum achievable rate against the number of relay antennas for $p_{\rm S}=10$ dB, $p_{\rm R}=10$ dB, $K=10$ and $b=1$.}
\label{fig3}
\end{figure}

In Fig. \ref{fig3}, we investigate the power scaling law of the mixed-ADC/DAC multipair mMIMO relaying system. The fraction of the number of high-resolution ADCs/DACs in the mixed-ADC/DAC architecture is $\kappa=0,1/2$ and 1, respectively. It can be seen that the exact and approximate expressions tend to a constant for $M \to \infty$, and a higher value of $\kappa$ increases the achievable rate, which agrees with Corollary \ref{corollary1}.


\begin{figure}[!t]
\centering
\includegraphics[scale=0.65]{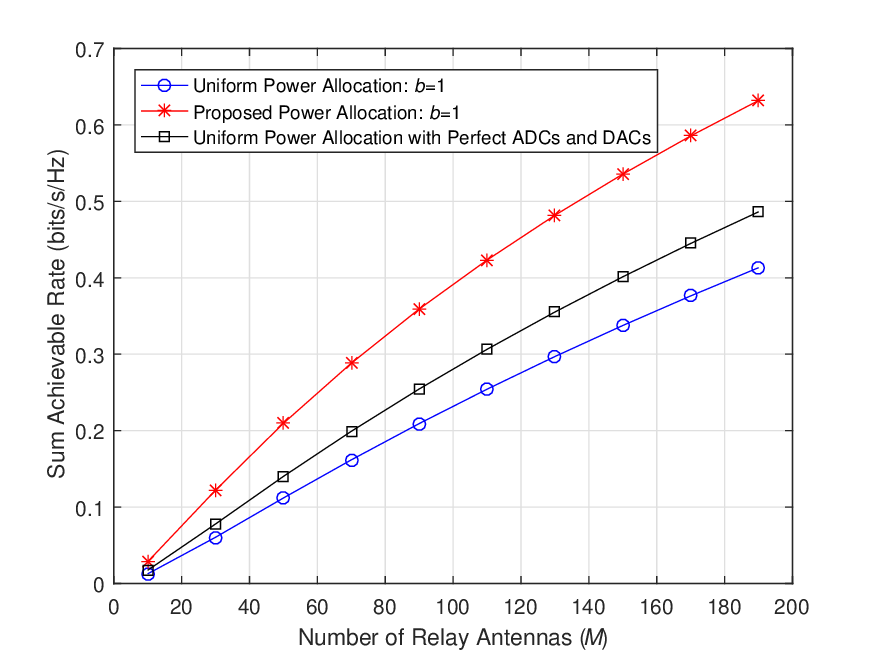}
\caption{Sum achievable rate against the number of relay antennas $M$ for $K=10$, $\kappa=1/2$ and $P_{\rm T}=10$ dB.}
\label{fig4}
\end{figure}

\subsection{Power Allocation}
We show the impact of the efficient power allocation scheme on the sum achievable rate in Fig. \ref{fig4}. The uniform power allocation scheme, i.e., $p_{\rm S}=\frac{P_{\rm T}}{2K}$ and $p_{\rm R}=\frac{P_{\rm T}}{K}$, is also investigated as a benchmark for comparison. It is clear that the proposed optimal power allocation scheme significantly boosts the sum rate compared with the one of the unquantized system with uniform power allocation. This important finding demonstrates the significance of adopting an efficient power allocation scheme in the mixed-ADC/DAC multipair mMIMO relaying system.

\subsection{Energy Efficiency}
Up to now, we have investigated the achievable rate of mixed-ADC/DAC multipair mMIMO relaying systems. As expected, under the same power allocation, the sum rate of unquantized system outperforms the one with mixed-resolution ADCs/DACs, at the cost of expensive hardware and power consumption. There should be a fundamental trade-off between the achievable rate and energy efficiency, and therefore, we also study the energy efficiency of mixed-ADC/DAC multipair mMIMO relaying systems.

According to \cite{zhang2017performance}, the energy efficiency can be defined as
\begin{equation}\label{eq78}
{\eta _{EE}} = \frac{{R \times B}}{{{P_{\rm total}}}}{{\rm bit} \mathord{\left/
 {\vphantom {{\rm bit} {\rm Joule}}} \right.
 \kern-\nulldelimiterspace} {\rm Joule}},
\end{equation}
where $R$ denotes the sum achievable rate, $B$ refers to the transmission bandwidth assumed to be 20 MHz, and $P_{\rm total}$ is the total power consumption. Combining \cite[Eq.  (43)]{zhang2017performance} and \cite[Eq.  (9)]{he2014energy}, $P_{\rm total}$ can be expressed as
\begin{align}\label{eq79}
&{P_{{\rm{total}}}}= M\left( {{P_{\rm mix}} + {P_{\rm filt}}} \right) + 2{P_{\rm syn}}+ {M_1}\left( {c{P_{\rm AGC}} + P_{\rm DAC}^L} \right)\notag \\
&+ M\left( {{P_{\rm LNA}} + {P_{\rm mix}} + {P_{\rm IFA}} + {P_{\rm filr}}} \right)+ {M_0}\left( {{P_{\rm AGC}} + P_{\rm ADC}^H} \right) \notag \\
&+ {M_1}\left( {c{P_{\rm AGC}} + P_{\rm ADC}^L} \right) + {M_0}\left( {{P_{\rm AGC}} + P_{\rm DAC}^H} \right),
\end{align}
where $P_{\rm mix}$, $P_{\rm filt}$, $P_{\rm syn}$, $P_{\rm LNA}$, $P_{\rm IFA}$, $P_{\rm filr}$, $P_{\rm AGC}$, $P_{\rm ADC}^H$, $P_{\rm ADC}^L$, $P_{\rm DAC}^H$ and $P_{\rm DAC}^L$ are the power consumption values for the mixer, the active filters at the transmitter side, the frequency synthesizer, low-noise amplifiers (LNA), the intermediate frequency amplifier (IFA), the active filters at the receiver side, the automatic gain control (AGC), high-resolution ADCs, low-resolution ADCs, high-resolution DACs and low-resolution DACs, respectively. In addition, $c$ denotes the flag related to quantization bits of low-resolution ADCs, {which} is given by
\begin{equation}\label{eq80}
c = \left\{ {\begin{array}{ll}
0,\quad b=1,\\
1,\quad b > 1.
\end{array}} \right.
\end{equation}
According to \cite{cui2005energy}, the power consumed in DACs and ADCs can be respectively expressed in terms of the number of quantization bits as
\begin{align}
{P_{\rm DAC}} &= \frac{1}{2}{V_{dd}}{I_0}\left( {{2^b} - 1} \right) + b{C_p}\left( {2B + {f_{cor}}} \right)V_{dd}^2,\label{eq81}\\
{P_{\rm ADC}} &= \frac{{3V_{dd}^2{L_{\min }}\left( {2B + {f_{cor}}} \right)}}{{{{10}^{ - 0.1525b + 4.838}}}},\label{eq82}
\end{align}
where $b$ denotes the quantization bits, {$B$ is the bandwidth of the original signal assumed to be {20 MHz}, $V_{dd}$ is the power supply of converter, $I_0$ is the unit current source corresponding to the {least significant bit} (LSB), $C_p$ is the parasitic capacitance of each switch in the converter, $L_{min}$ is the minimum channel length for the given CMOS technology, $f_{cor}$ is the corner frequency of the $1/f$ noise and all those parameters are specifically  defined in \cite{cui2005energy}. (\ref{eq81}) {holds} for binary-weighted current-steering DACs \cite{cui2005energy} and (\ref{eq82}) is established for the complete class of CMOS Nyquist-rate {high speed} ADCs \cite{lauwers2002power}.} In numerical examples, we {consider} the following classical values: $P_{\rm mix}=30.3$ mW, $P_{\rm filt}=P_{\rm filr}=2.5$ mW, $P_{\rm syn}=50.0$ mW, $P_{\rm LNA}=20$ mW, $P_{\rm IFA}=3$ mW and $P_{\rm AGC}=2$ mW as in \cite{zhang2017performance} and \cite{he2014energy}, {and the power consumption values of other various circuit blocks have been discussed in \cite{cui2005energy}}.

\begin{figure}[!t]
\centering
\includegraphics[scale=0.65]{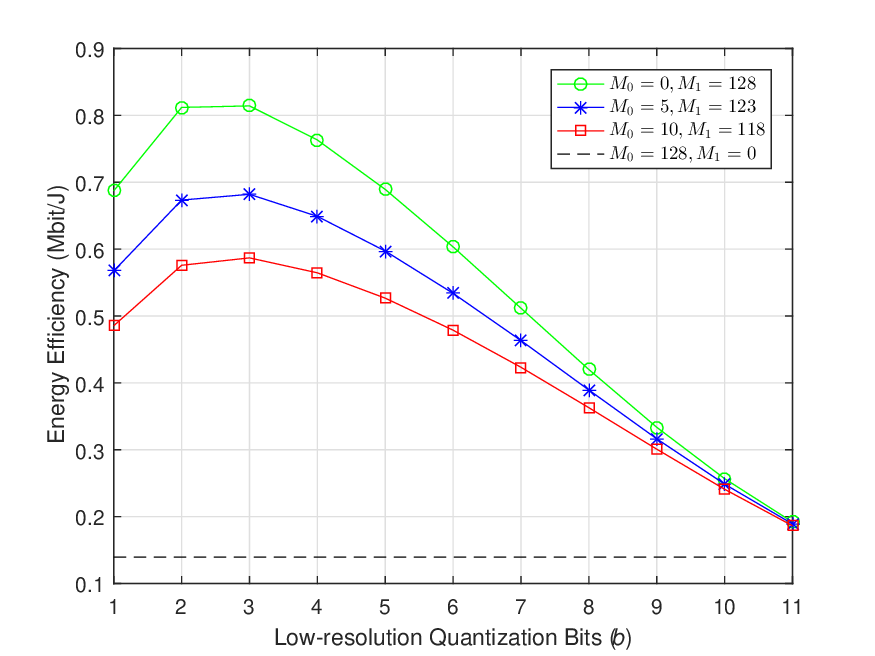}
\caption{Energy efficiency of multipair mMIMO relaying systems with mixed-resolution ADCs/DACs against the number of low-resolution quantization bits $b$  for $p_{\rm S}=10$ dB, $p_{\rm R}=10$ dB and $K=10$.}
\label{fig5}
\end{figure}

The energy efficiency of mixed-ADC/DAC multipair mMIMO relaying systems against the quantization bits is illustrated in Fig. \ref{fig5}. It is clear from the figure that the relay adopting pure low-resolution ADCs/DACs attains the best energy efficiency. That means the energy efficiency increases with the proportion of the number of low-resolution ADCs/DACs ($1-\kappa$) in the mixed-ADC/DAC architecture. {{Although the pure low-resolution ADC/DAC architecture can achieve better energy efficiency than the mixed-ADC/DAC architecture, as shown in \ref{fig1}, the spectral efficiency of the low-resolution ADC/DAC architecture is much lower than that of the mixed-ADC/DAC architecture. Moreover, the channel estimation in the mixed-ADC/DAC architecture is more tractable than that in the low-resolution ADC/DAC architecture due to the use of partial high-resolution ADCs}.}  {Moreover, the pure low-resolution ADCs/DACs has low spectrum efficiency.} Fig. \ref{fig5} indicates that we can achieve a better spectral efficiency by reducing the burden of power consumption considerably by using the mixed-ADC/DAC architecture.

\section{Conclusions}\label{se:conclusion}
In this paper, we investigated the achievable rate of a multipair mMIMO relaying system with mixed-resolution ADCs and DACs at the relay. Both exact and approximate closed-form expressions for the achievable rate were derived. Then, we proved that the transmit power of each users can be scaled down as ${1/ M}$ for the considered system. Despite the rate loss due to the use of low-resolution ADCs and DACs, employing massive antenna arrays still enables high achievable rate and large power saving. Furthermore, we proposed an efficient power allocation scheme, which can compensate for the rate degradation caused by low-resolution ADCs and DACs. Finally, the energy efficiency was investigated, and {showed} that the mixed-ADC/DAC architecture can attain a considerable rate and energy efficiency simultaneously, which is promising for practical mMIMO relaying systems.

\appendices
\section{Proof of Lemma \ref{lemma1}}\label{ap1}
For ${{\mathbb{E}}\left\{ {{{\left\| {{{{\boldsymbol {\rm{\tilde x}}}}_{\rm{R}}}} \right\|}^2}} \right\}} = {{\mathbb{E}}\left\{ {{{\left\| {{{{\boldsymbol {\rm{\tilde x}}}}_{\rm{R0}}}} \right\|}^2}} \right\}}+{{\mathbb{E}}\left\{ {{{\left\| {{{{\boldsymbol {\rm{\tilde x}}}}_{\rm{R1}}}} \right\|}^2}} \right\}}$, we first

\subsubsection{\rm{Calculate ${{\mathbb{E}}\left\{{{{\left\| {{{{\boldsymbol {\rm{\tilde x}}}}_{\rm{R0}}}} \right\|}^2}} \right\}}$ }}
\begin{equation}\label{eq83}
\begin{aligned}
{{\boldsymbol{\rm{R}}}_{{{{\boldsymbol{\rm{\tilde x}}}}_{{\rm{R0}}}}{{{\boldsymbol{\rm{\tilde x}}}}_{{\rm{R0}}}}}}&= {\boldsymbol{\rm{G}}}_{{\rm{RD0}}}^*{\boldsymbol{\rm{G}}}_{{\rm{SR0}}}^H{{\boldsymbol{\rm{R}}}_{{{{\boldsymbol{\rm{\tilde y}}}}_{{\rm{R0}}}}{{{\boldsymbol{\rm{\tilde y}}}}_{{\rm{R0}}}}}}{{\boldsymbol{\rm{G}}}_{{\rm{SR0}}}}{\boldsymbol{\rm{G}}}_{{\rm{RD0}}}^T \\
&+ {\boldsymbol{\rm{G}}}_{{\rm{RD0}}}^*{\boldsymbol{\rm{G}}}_{{\rm{SR0}}}^H{{\boldsymbol{\rm{R}}}_{{{{\boldsymbol{\rm{\tilde y}}}}_{{\rm{R0}}}}{{{\boldsymbol{\rm{\tilde y}}}}_{{\rm{R1}}}}}}{{\boldsymbol{\rm{G}}}_{{\rm{SR1}}}}{\boldsymbol{\rm{G}}}_{{\rm{RD0}}}^T \\
&+ {\boldsymbol{\rm{G}}}_{{\rm{RD0}}}^*{\boldsymbol{\rm{G}}}_{{\rm{SR1}}}^H{{\boldsymbol{\rm{R}}}_{{{{\boldsymbol{\rm{\tilde y}}}}_{{\rm{R1}}}}{{{\boldsymbol{\rm{\tilde y}}}}_{{\rm{R0}}}}}}{{\boldsymbol{\rm{G}}}_{{\rm{SR0}}}}{\boldsymbol{\rm{G}}}_{{\rm{RD0}}}^T\\
&+ {\boldsymbol{\rm{G}}}_{{\rm{RD0}}}^*{\boldsymbol{\rm{G}}}_{{\rm{SR1}}}^H{{\boldsymbol{\rm{R}}}_{{{{\boldsymbol{\rm{\tilde y}}}}_{{\rm{R1}}}}{{{\boldsymbol{\rm{\tilde y}}}}_{{\rm{R1}}}}}}{{\boldsymbol{\rm{G}}}_{{\rm{SR1}}}}{\boldsymbol{\rm{G}}}_{{\rm{RD0}}}^T\\
&={\boldsymbol{\rm Q}}_1+{\boldsymbol{\rm Q}}_2+{\boldsymbol{\rm Q}}_3+{\boldsymbol{\rm Q}}_4,
\end{aligned}
\end{equation}
where ${{\boldsymbol{\rm{R}}}_{{{{\boldsymbol{\rm{\tilde y}}}}_{{\rm{R0}}}}{{{\boldsymbol{\rm{\tilde y}}}}_{{\rm{R0}}}}}}$, ${{\boldsymbol{\rm{R}}}_{{{{\boldsymbol{\rm{\tilde y}}}}_{{\rm{R0}}}}{{{\boldsymbol{\rm{\tilde y}}}}_{{\rm{R1}}}}}}$, ${{\boldsymbol{\rm{R}}}_{{{{\boldsymbol{\rm{\tilde y}}}}_{{\rm{R1}}}}{{{\boldsymbol{\rm{\tilde y}}}}_{{\rm{R0}}}}}}$
and ${{\boldsymbol{\rm{R}}}_{{{{\boldsymbol{\rm{\tilde y}}}}_{{\rm{R1}}}}{{{\boldsymbol{\rm{\tilde y}}}}_{{\rm{R1}}}}}}$ are separately given by (\ref{eq18}) to (\ref{eq21}), and ${\boldsymbol{\rm Q}}_1$, ${\boldsymbol{\rm Q}}_2$ ,${\boldsymbol{\rm Q}}_3$ and ${\boldsymbol{\rm Q}}_4$ are defined as the four parts of  ${{\boldsymbol{\rm{R}}}_{{{{\boldsymbol{\rm{\tilde x}}}}_{{\rm{R0}}}}{{{\boldsymbol{\rm{\tilde x}}}}_{{\rm{R0}}}}}}$, respectively.
\begin{align}\label{eq84}
\mathbb{E}\left\{{{\boldsymbol{\rm Q}}_1}\right\}&= \mathbb{E}\left\{{{\boldsymbol{\rm{G}}}_{{\rm{RD0}}}^*{\boldsymbol{\rm{G}}}_{{\rm{SR0}}}^H{\boldsymbol {\rm G}_{\rm SR0}{{\boldsymbol{\rm{P}}}_{S}}{\boldsymbol {\rm G}_{\rm SR0}^H}}{{\boldsymbol{\rm{G}}}_{{\rm{SR0}}}}{\boldsymbol{\rm{G}}}_{{\rm{RD0}}}^T}\right\} \notag \\
&+\mathbb{E}\left\{{{\boldsymbol{\rm{G}}}_{{\rm{RD0}}}^*{\boldsymbol{\rm{G}}}_{{\rm{SR0}}}^H
{{\boldsymbol{\rm{G}}}_{{\rm{SR0}}}}{\boldsymbol{\rm{G}}}_{{\rm{RD0}}}^T}\right\}.
\end{align}

Using  $\mathbb{E}\left\{{{{{\left\| {{\boldsymbol{\rm g}_{{\rm {SR0}},m}}} \right\|}^4}}}\right\}=M_0\left({M_0+1}\right)\beta_{{\rm SR},m}^2$, we have
\begin{align}\label{eq86}
{\mathbb{E}\left\{{{\boldsymbol{\rm{G}}}_{{\rm{RD0}}}^*{\boldsymbol{\rm{G}}}_{{\rm{SR0}}}^H{{\boldsymbol{\rm{G}}}_{{\rm{SR0}}}}{\boldsymbol{\rm{G}}}_{{\rm{RD0}}}^T}\right\}} = M_0\sum\limits_{m = 1}^K {{\beta _{{\rm{SR}},m}}{\beta _{{\rm{RD}},m}}} {{\boldsymbol{\rm I}}_{M0}},
\end{align}
\begin{align}\label{eq85}
&\mathbb{E}\left\{{{\boldsymbol{\rm{G}}}_{{\rm{RD0}}}^*{\boldsymbol{\rm{G}}}_{{\rm{SR0}}}^H{\boldsymbol {\rm G}_{\rm SR0}{{\boldsymbol{\rm{P}}}_{S}}{\boldsymbol {\rm G}_{\rm SR0}^H}}{{\boldsymbol{\rm{G}}}_{{\rm{SR0}}}}{\boldsymbol{\rm{G}}}_{{\rm{RD0}}}^T}\right\}\notag \\
&=M_0\sum\limits_{m = 1}^K {{\beta _{{\rm{SR}},m}}{\beta _{{\rm{RD}},m}}\left({\sum\limits_{i = 1}^K {{p_{{\rm S},i}}{\beta _{{\rm{SR}},i}}} \!+\!M_0{p_{{\rm S},m}}{\beta _{{\rm{SR}},m}}} \right)} {{\boldsymbol{\rm I}}_{M0}}.
\end{align}
Then substituting (\ref{eq85}) and (\ref{eq86}) into (\ref{eq84}), we directly obtain
\begin{equation}\label{eq87}
\begin{aligned}
\mathbb{E}\left\{{{\boldsymbol{\rm Q}}_1}\right\}&= M_0\sum\limits_{m = 1}^K {{\beta _{{\rm{SR}},m}}{\beta _{{\rm{RD}},m}}}\\
&\times \left({\sum\limits_{i = 1}^K {{p_{{\rm S},i}}{\beta _{{\rm{SR}},i}}} +M_0{p_{{\rm S},m}}{\beta _{{\rm{SR}},m}}+1} \right) {{\boldsymbol{\rm I}}_{M0}}.
\end{aligned}
\end{equation}
Similar to the computation of ${\boldsymbol{\rm Q}}_1$, the expectation of ${\boldsymbol{\rm Q}}_2$, ${\boldsymbol{\rm Q}}_3$ and ${\boldsymbol{\rm Q}}_4$ can be derived respectively as
\begin{align}
\mathbb{E}\left\{{{\boldsymbol{\rm Q}}_2}\right\}&= \alpha M_0 M_1\sum\limits_{m = 1}^K {{p_{{\rm S},m}}{\beta^2 _{{\rm{SR}},m}}{\beta _{{\rm{RD}},m}}}{{\boldsymbol{\rm I}}_{M0}},\label{eq88}\\
\mathbb{E}\left\{{{\boldsymbol{\rm Q}}_3}\right\}&= \alpha M_0 M_1\sum\limits_{m = 1}^K {{p_{{\rm S},m}}{\beta^2 _{{\rm{SR}},m}}{\beta _{{\rm{RD}},m}}}{{\boldsymbol{\rm I}}_{M0}}. \label{eq89}\\
\mathbb{E}\left\{{{\boldsymbol{\rm Q}}_4}\right\}&= \alpha^2M_1\sum\limits_{m = 1}^K {{\beta _{{\rm{SR}},m}}{\beta _{{\rm{RD}},m}}} \notag \\
& \times \left( \sum\limits_{i = 1}^K {{p_{{\rm S},i}}{\beta _{{\rm{SR}},i}}}+M_1{p_{{\rm S},m}}{\beta _{{\rm{SR}},m}}+1  \right) {{\boldsymbol{\rm I}}_{M0}}\notag\\
&+\alpha\rho\mathbb{E}\left\{{{\boldsymbol{\rm{G}}}_{{\rm{RD0}}}^*{\boldsymbol{\rm{G}}}_{{\rm{SR1}}}^H\boldsymbol {\rm R}_{{\boldsymbol{\rm{n}}}_{{{\rm{q}}_{\rm{a}}}}{\boldsymbol{\rm{n}}}_{{{\rm{q}}_{\rm{a}}}}}
{{\boldsymbol{\rm{G}}}_{{\rm{SR1}}}}{\boldsymbol{\rm{G}}}_{{\rm{RD0}}}^T}\right\}.\label{eq90}
\end{align}
{{where  }}
\begin{align}\label{eq91}
&\mathbb{E}\left\{{{\boldsymbol{\rm{G}}}_{{\rm{RD0}}}^*{\boldsymbol{\rm{G}}}_{{\rm{SR1}}}^H\boldsymbol {\rm R}_{{\boldsymbol{\rm{n}}}_{{{\rm{q}}_{\rm{a}}}}{\boldsymbol{\rm{n}}}_{{{\rm{q}}_{\rm{a}}}}}{{\boldsymbol{\rm{G}}}_{{\rm{SR1}}}}{\boldsymbol{\rm{G}}}_{{\rm{RD0}}}^T}\right\}\notag\\
&=\mathbb{E}\left\{{{\boldsymbol{\rm{G}}}_{{\rm{RD0}}}^*\mathbb{E}\left[{\boldsymbol{\rm{G}}}_{{\rm{SR1}}}^H{\rm{diag}}\left( {\boldsymbol {\rm G}_{\rm SR1}{{\boldsymbol{\rm{P}}}_{S}}{\boldsymbol {\rm G}_{\rm SR1}^H}} \right){{\boldsymbol{\rm{G}}}_{{\rm{SR1}}}}\right]{\boldsymbol{\rm{G}}}_{{\rm{RD0}}}^T}\right\}\notag\\
&+M_1\sum\limits_{m = 1}^K {{\beta _{{\rm{SR}},m}}{\beta _{{\rm{RD}},m}}} {{\boldsymbol{\rm I}}_{M0}}.
\end{align}
The expectation of the diagonal term can be decomposed as
\begin{equation}\label{eq92}
\begin{aligned}
&\mathbb{E}\left\{{\boldsymbol{\rm{g}}}_{{\rm{SR1}},i}^H{\rm{diag}}\left( {\boldsymbol {\rm G}_{\rm SR1}{{\boldsymbol{\rm{P}}}_{S}}{\boldsymbol {\rm G}_{\rm SR1}^H}} \right){{\boldsymbol{\rm{g}}}_{{\rm{SR1}},j}}\right\}\\
=&\left\{ {\begin{array}{ll}
0,& i \ne j,\\
{\begin{aligned}&\sum\limits_{m = 1}^{{M_1}} {{p_{{S},n}}{\mathbb{E}}\left\{ {{{\left| {{g_{{\rm{SR1}},mn}}} \right|}^4}} \right\}}  \\
& \!+\!
 \sum\limits_{m \!=\! 1}^{{M_1}} {\sum\limits_{i \ne n}^K {{p_{{S},i}}{\mathbb{E}}\left\{ {{{\left| {{g_{{\rm{SR1}},{{mn}}}}} \right|}^2}} \right\}{\mathbb{E}}\left\{ {{{\left| {{g_{{\rm{SR1}},{{mi}}}}} \right|}^2}} \right\}} }\end{aligned}} ,&i\!=\!j\!=\!n.
\end{array}} \right.
\end{aligned}
\end{equation}
Applying the fact that ${\mathbb{E}}\left\{ {{{\left| {{g_{{\rm{SR1}},{{mn}}}}} \right|}^4}}\right\}=2\beta^2_{{\rm{SR}},n}$, (\ref{eq92}) can be expressed as
\begin{equation}\label{eq93}
\begin{aligned}
&\mathbb{E}\left\{{\boldsymbol{\rm{g}}}_{{\rm{SR1}},i}^H{\rm{diag}}\left( {\boldsymbol {\rm G}_{\rm SR1}{{\boldsymbol{\rm{P}}}_{S}}{\boldsymbol {\rm G}_{\rm SR1}^H}} \right){{\boldsymbol{\rm{g}}}_{{\rm{SR1}},j}}\right\}\\
=&\left\{ {\begin{array}{ll}
0,& i \ne j,\\
{M_1}{\beta _{{\rm{SR}},n}}\left( {\sum\limits_{i = 1}^K {{p_{{S},i}}{\beta _{{\rm{SR}},i}}}  + {p_{{S},n}}{\beta _{{\rm{SR}},n}}} \right),&i=j=n.
\end{array}} \right.
\end{aligned}
\end{equation}
Substituting (\ref{eq93}) into (\ref{eq91}), we can obtain
\begin{align}\label{eq94}
&\mathbb{E}\left\{{{\boldsymbol{\rm{G}}}_{{\rm{RD0}}}^*{\boldsymbol{\rm{G}}}_{{\rm{SR1}}}^H\boldsymbol {\rm R}_{{\boldsymbol{\rm{n}}}_{{{\rm{q}}_{\rm{a}}}}{\boldsymbol{\rm{n}}}_{{{\rm{q}}_{\rm{a}}}}}{{\boldsymbol{\rm{G}}}_{{\rm{SR1}}}}{\boldsymbol{\rm{G}}}_{{\rm{RD0}}}^T}\right\}\notag \\
&= M_1\sum\limits_{m = 1}^K {{\beta _{{\rm{SR}},m}}{\beta _{{\rm{RD}},m}}}\left({\sum\limits_{i = 1}^K {{p_{{\rm S},i}}{\beta _{{\rm{SR}},i}}} \!+\! {p_{{\rm S},m}}{\beta _{{\rm{SR}},m}}\!+\!1} \right) {{\boldsymbol{\rm I}}_{M0}}.
\end{align}
With the help of (\ref{eq90}) and (\ref{eq94}), we derive $\mathbb{E}\left\{{{\boldsymbol{\rm Q}}_4}\right\}$ as
\begin{equation}\label{eq95}
\begin{aligned}
&\mathbb{E}\left\{{{\boldsymbol{\rm Q}}_4}\right\} =\alpha M_1\sum\limits_{m = 1}^K {{\beta _{{\rm{SR}},m}}{\beta _{{\rm{RD}},m}}}\\
&\times\left({\sum\limits_{i = 1}^K {{p_{{\rm S},i}}{\beta _{{\rm{SR}},i}}} \!+\!\alpha M_1{p_{{\rm S},m}}{\beta _{{\rm{SR}},m}}\!+\!\rho {p_{{\rm S},m}}{\beta _{{\rm{SR}},m}}\!+\!1} \right) {{\boldsymbol{\rm I}}_{M0}}.
\end{aligned}
\end{equation}
Combining (\ref{eq87}), (\ref{eq88}), (\ref{eq89}) and (\ref{eq95}), we can derive
\begin{equation}\label{eq96}
\mathbb{E}\left\{{{\boldsymbol{\rm{R}}}_{{{{\boldsymbol{\rm{\tilde x}}}}_{{\rm{R0}}}}{{{\boldsymbol{\rm{\tilde x}}}}_{{\rm{R0}}}}}}\right\}=\mathbb{E}\left\{{{\boldsymbol{\rm{R}}}_{{{{\boldsymbol{\rm{x}}}}_{{\rm{R0}}}}{{{\boldsymbol{\rm{x}}}}_{{\rm{R0}}}}}}\right\}=\mu  {\boldsymbol{\rm I}}_{M0}.
\end{equation}
\begin{equation}\label{eq97}
{{\mathbb{E}}\left\{ {{{\left\| {{{{\boldsymbol {\rm{\tilde x}}}}_{\rm{R0}}}} \right\|}^2}} \right\}}={{\mathbb{E}}\left\{ {{{\left\| {{{{\boldsymbol {\rm{x}}}}_{\rm{R0}}}} \right\|}^2}} \right\}}=\mu M_0.
\end{equation}

\subsubsection{\rm{Calculate ${{\mathbb{E}}\left\{ {{{\left\| {{{{\boldsymbol {\rm{\tilde x}}}}_{\rm{R1}}}} \right\|}^2}} \right\}}$}}
\begin{equation}\label{eq98}
\begin{aligned}
{{\mathbb{E}}\left\{ {{{\left\| {{{{\boldsymbol {\rm{\tilde x}}}}_{\rm{R1}}}} \right\|}^2}} \right\}} =\alpha^2{{\mathbb{E}}\left\{ {{{\left\| {{{{\boldsymbol {\rm{ x}}}}_{\rm{R1}}}} \right\|}^2}} \right\}}+\mathbb{E}\left\{{{{\boldsymbol{\rm{n}}}^H_{{{\rm{q}}_{D}}}}{{\boldsymbol{\rm{n}}}_{{{\rm{q}}_{D}}}}}\right\}.
\end{aligned}
\end{equation}
Similar to the calculation of (\ref{eq96}) and (\ref{eq97}), $\mathbb{E}\left\{{{\boldsymbol{\rm{R}}}_{{{{\boldsymbol{\rm{x}}}}_{{\rm{R0}}}}{{{\boldsymbol{\rm{x}}}}_{{\rm{R0}}}}}}\right\}$ and ${{\mathbb{E}}\left\{ {{{\left\| {{{{\boldsymbol {\rm{ x}}}}_{\rm{R1}}}} \right\|}^2}} \right\}}$ can be respectively expressed as
\begin{equation}\label{eq99}
\mathbb{E}\left\{{{\boldsymbol{\rm{R}}}_{{{{\boldsymbol{\rm{x}}}}_{{\rm{R1}}}}{{{\boldsymbol{\rm{x}}}}_{{\rm{R1}}}}}}\right\}=\mu  {\boldsymbol{\rm I}}_{M1},
\end{equation}
\begin{equation}\label{eq100}
{{\mathbb{E}}\left\{ {{{\left\| {{{{\boldsymbol {\rm{x}}}}_{\rm{R1}}}} \right\|}^2}} \right\}}=\mu M_1.
\end{equation}
As for $\mathbb{E}\left\{{{{\boldsymbol{\rm{n}}}^H_{{{\rm{q}}_{D}}}}{{\boldsymbol{\rm{n}}}_{{{\rm{q}}_{D}}}}}\right\}$, considering (\ref{eq16}) and (\ref{eq99}), we can derive $\boldsymbol {\rm R}_{{\boldsymbol{\rm{n}}}_{{{\rm{q}}_{D}}}}$ as
\begin{equation}\label{eq101}
\mathbb{E}\left\{\boldsymbol {\rm R}_{{\boldsymbol{\rm{n}}}_{{{\rm{q}}_{D}}}}\right\}=\alpha\rho\mathbb{E}\left\{{{\boldsymbol{\rm{R}}}_{{{{\boldsymbol{\rm{x}}}}_{{\rm{R1}}}}{{{\boldsymbol{\rm{x}}}}_{{\rm{R1}}}}}}\right\}=\alpha\rho\mu{\boldsymbol{\rm I}}_{M1}.
\end{equation}
Hence,
\begin{equation}\label{eq102}
\mathbb{E}\left\{{{{\boldsymbol{\rm{n}}}^H_{{{\rm{q}}_{D}}}}{{\boldsymbol{\rm{n}}}_{{{\rm{q}}_{D}}}}}\right\}=\alpha\rho\mu M_1.
\end{equation}
Substituting (\ref{eq100}) and (\ref{eq102}) into (\ref{eq98}), we can directly obtain
\begin{equation}\label{eq103}
{{\mathbb{E}}\left\{ {{{\left\| {{{{\boldsymbol {\rm{\tilde x}}}}_{\rm{R1}}}} \right\|}^2}} \right\}}=\mu \alpha M_1.
\end{equation}
Therefore, ${{\mathbb{E}}\left\{ {{{\left\| {{{{\boldsymbol {\rm{\tilde x}}}}_{\rm{R}}}} \right\|}^2}} \right\}} = {{\mathbb{E}}\left\{ {{{\left\| {{{{\boldsymbol {\rm{\tilde x}}}}_{\rm{R0}}}} \right\|}^2}} \right\}}+{{\mathbb{E}}\left\{ {{{\left\| {{{{\boldsymbol {\rm{\tilde x}}}}_{\rm{R1}}}} \right\|}^2}} \right\}}=\mu\left({M_0+\alpha M_1}\right)$. The proof is concluded.

\section{Proof of Theorem \ref{theorem1}}\label{ap2}
The argument of the log function in the right-hand side of (\ref{eq29}) consists of six terms: 1) desired signal power $A_k$; 2) estimation error $B_k$; 3) inter-pair interference $C_k$; 4) noise at the relay $D_k$ and $E_k$; 5) quantization noise of ADCs $F_k$; 6) quantization noise of DACs $G_k$.
\setcounter{subsubsection}{0}
\subsubsection{\rm{Compute $A_k$}}
Since
\begin{align}\label{eq104}
{{\mathbb{E}}\left\{ {T_{k,k}} \right\}}= \gamma\left({M_0+\alpha M_1}\right)^2\beta_{{\rm SR},k}\beta_{{\rm RD},k},
\end{align}
we have
\begin{equation}\label{eq105}
{A_k}={p_{{S},k}}{\gamma ^2}{\left( {{M_0} + \alpha {M_1}} \right)^4}\beta _{{\rm{SR}},k}^2\beta _{{\rm{RD}},k}^2.
\end{equation}

\subsubsection{\rm{Compute $B_k$}}
\begin{equation}\label{eq106}
{B_k}={p_{{S},k}}{\rm Var}\left({T_{k,k}}\right)={p_{{S},k}}{\mathbb{E}}\left\{ {\left|T_{k,k}\right|^2}\right\}-A_k,
\end{equation}
We define $t_1$ to $t_{10}$ as the decomposed terms of ${\mathbb{E}}\left\{{T_{k,i}T^H_{k,i}}\right\}$ in order. Note that the undefined terms in $t_k$ mean that they are included in the expressions if and only if $i=k$.
\begin{align}
t_1&={\mathbb{E}}\left\{{{\boldsymbol{\rm g}_{{\rm{RD0}},k}^T{\boldsymbol{\rm G}}_{{\rm{RD0}}}^*{\boldsymbol{\rm G}}_{{\rm{SR0}}}^H{\boldsymbol{\rm g}_{{\rm{SR0}},i}}}{\boldsymbol{\rm g}^H_{{\rm{SR0}},i}}{\boldsymbol{\rm G}}_{{\rm{SR0}}}{\boldsymbol{\rm G}}_{{\rm{RD0}}}^T{\boldsymbol{\rm g}^*_{{\rm{RD0}},k}}}\right\}\notag\\
&=M^2_0\beta_{{\rm SR},i}\beta_{{\rm RD},k}({M_0\beta_{{\rm SR},k}\beta_{{\rm RD},k}\!+\!\sum\limits_{m = 1}^K {{\beta _{{\rm{SR}},m}}{\beta _{{\rm{RD}},m}}}} \notag \\
&\!+\!M_0\beta_{{\rm SR},i}\beta_{{\rm RD},i})+{M^4_0\beta^2_{{\rm SR},k}\beta^2_{{\rm RD},k}},\label{eq108}\\
t_2&={\mathbb{E}}\left\{{\alpha^2{\boldsymbol{\rm g}_{{\rm{RD0}},k}^T{\boldsymbol{\rm G}}_{{\rm{RD0}}}^*{\boldsymbol{\rm G}}_{{\rm{SR1}}}^H{\boldsymbol{\rm g}_{{\rm{SR1}},i}}}{\boldsymbol{\rm g}^H_{{\rm{SR1}},i}}{\boldsymbol{\rm G}}_{{\rm{SR1}}}{\boldsymbol{\rm G}}_{{\rm{RD0}}}^T{\boldsymbol{\rm g}^*_{{\rm{RD0}},k}}}\right\}\notag\\
&=\alpha^2M_0M_1\beta_{{\rm SR},i}\beta_{{\rm RD},k}({M_0\beta_{{\rm SR},k}\beta_{{\rm RD},k}\!+\!\sum\limits_{m = 1}^K {{\beta _{{\rm{SR}},m}}{\beta _{{\rm{RD}},m}}}} \notag\\
&\!+\!M_1\beta_{{\rm SR},i}\beta_{{\rm RD},i}) +{\alpha^2M^2_0M^2_1\beta^2_{{\rm SR},k}\beta^2_{{\rm RD},k}},\label{eq109}\\
t_3&={\mathbb{E}}\left\{{\alpha^2{\boldsymbol{\rm g}_{{\rm{RD1}},k}^T{\boldsymbol{\rm G}}_{{\rm{RD1}}}^*{\boldsymbol{\rm G}}_{{\rm{SR0}}}^H{\boldsymbol{\rm g}_{{\rm{SR0}},i}}}{\boldsymbol{\rm g}^H_{{\rm{SR0}},i}}{\boldsymbol{\rm G}}_{{\rm{SR0}}}{\boldsymbol{\rm G}}_{{\rm{RD1}}}^T{\boldsymbol{\rm g}^*_{{\rm{RD1}},k}}}\right\}\notag\\
&=\alpha^2M_0M_1\beta_{{\rm SR},i}\beta_{{\rm RD},k}({M_1\beta_{{\rm SR},k}\beta_{{\rm RD},k}+\sum\limits_{m = 1}^K {{\beta _{{\rm{SR}},m}}{\beta _{{\rm{RD}},m}}}}\notag \\
& +M_0\beta_{{\rm SR},i}\beta_{{\rm RD},i})+{\alpha^2M^2_0M^2_1\beta^2_{{\rm SR},k}\beta^2_{{\rm RD},k}},\label{eq110}\\
t_4&={\mathbb{E}}\left\{{\alpha^4{\boldsymbol{\rm g}_{{\rm{RD1}},k}^T{\boldsymbol{\rm G}}_{{\rm{RD1}}}^*{\boldsymbol{\rm G}}_{{\rm{SR1}}}^H{\boldsymbol{\rm g}_{{\rm{SR1}},i}}}{\boldsymbol{\rm g}^H_{{\rm{SR1}},i}}{\boldsymbol{\rm G}}_{{\rm{SR1}}}{\boldsymbol{\rm G}}_{{\rm{RD1}}}^T{\boldsymbol{\rm g}^*_{{\rm{RD1}},k}}}\right\}\notag\\
&=\alpha^4M^2_1\beta_{{\rm SR},i}\beta_{{\rm RD},k}\Big({M_1\beta_{{\rm SR},k}\beta_{{\rm RD},k}+\sum\limits_{m = 1}^K {{\beta _{{\rm{SR}},m}}{\beta _{{\rm{RD}},m}}}} \notag \\
&+M_1\beta_{{\rm SR},i}\beta_{{\rm RD},i}\Big) +{\alpha^4M^4_1\beta^2_{{\rm SR},k}\beta^2_{{\rm RD},k}},\label{eq111}
\end{align}
\begin{align}
t_5&={\mathbb{E}}\left\{{\alpha{\boldsymbol{\rm g}_{{\rm{RD0}},k}^T{\boldsymbol{\rm G}}_{{\rm{RD0}}}^*{\boldsymbol{\rm G}}_{{\rm{SR0}}}^H{\boldsymbol{\rm g}_{{\rm{SR0}},i}}}{\boldsymbol{\rm g}^H_{{\rm{SR1}},i}}{\boldsymbol{\rm G}}_{{\rm{SR1}}}{\boldsymbol{\rm G}}_{{\rm{RD0}}}^T{\boldsymbol{\rm g}^*_{{\rm{RD0}},k}}}\right\}\notag \\
&=\alpha M^2_0M_1\beta^2_{{\rm SR},i}\beta_{{\rm RD},k}\left({\beta_{{\rm RD},i}+{M_0\beta_{{\rm RD},k}}}\right),\label{eq112}\\
t_6&={\mathbb{E}}\left\{{\alpha{\boldsymbol{\rm g}_{{\rm{RD0}},k}^T{\boldsymbol{\rm G}}_{{\rm{RD0}}}^*{\boldsymbol{\rm G}}_{{\rm{SR0}}}^H{\boldsymbol{\rm g}_{{\rm{SR0}},i}}}{\boldsymbol{\rm g}^H_{{\rm{SR0}},i}}{\boldsymbol{\rm G}}_{{\rm{SR0}}}{\boldsymbol{\rm G}}_{{\rm{RD1}}}^T{\boldsymbol{\rm g}^*_{{\rm{RD1}},k}}}\right\}\notag \\
&=\alpha M^2_0M_1\beta_{{\rm SR},k}\beta^2_{{\rm RD},k}({\beta_{{\rm SR},i}+{M_0\beta_{{\rm SR},k}}}),\label{eq113}\\
t_7&={\mathbb{E}}\left\{{\alpha^2{\boldsymbol{\rm g}_{{\rm{RD0}},k}^T{\boldsymbol{\rm G}}_{{\rm{RD0}}}^*{\boldsymbol{\rm G}}_{{\rm{SR0}}}^H{\boldsymbol{\rm g}_{{\rm{SR0}},i}}}{\boldsymbol{\rm g}^H_{{\rm{SR1}},i}}{\boldsymbol{\rm G}}_{{\rm{SR1}}}{\boldsymbol{\rm G}}_{{\rm{RD1}}}^T{\boldsymbol{\rm g}^*_{{\rm{RD1}},k}}}\right\}\notag \\
&={\alpha^2M^2_0M^2_1\beta^2_{{\rm SR},k}\beta^2_{{\rm RD},k}},\label{eq114}\\
t_8&={\mathbb{E}}\left\{{\alpha^2{\boldsymbol{\rm g}_{{\rm{RD0}},k}^T{\boldsymbol{\rm G}}_{{\rm{RD0}}}^*{\boldsymbol{\rm G}}_{{\rm{SR1}}}^H{\boldsymbol{\rm g}_{{\rm{SR1}},i}}}{\boldsymbol{\rm g}^H_{{\rm{SR0}},i}}{\boldsymbol{\rm G}}_{{\rm{SR0}}}{\boldsymbol{\rm G}}_{{\rm{RD1}}}^T{\boldsymbol{\rm g}^*_{{\rm{RD1}},k}}}\right\}\notag \\
&={\alpha^2M^2_0M^2_1\beta^2_{{\rm SR},k}\beta^2_{{\rm RD},k}},\label{eq115}\\
t_9&={\mathbb{E}}\left\{{\alpha^3{\boldsymbol{\rm g}_{{\rm{RD0}},k}^T{\boldsymbol{\rm G}}_{{\rm{RD0}}}^*{\boldsymbol{\rm G}}_{{\rm{SR1}}}^H{\boldsymbol{\rm g}_{{\rm{SR1}},i}}}{\boldsymbol{\rm g}^H_{{\rm{SR1}},i}}{\boldsymbol{\rm G}}_{{\rm{SR1}}}{\boldsymbol{\rm G}}_{{\rm{RD1}}}^T{\boldsymbol{\rm g}^*_{{\rm{RD1}},k}}}\right\}\notag \\
&=\alpha^3 M_0M^2_1\beta_{{\rm SR},k}\beta^2_{{\rm RD},k}({\beta_{{\rm SR},i}+{M_1\beta_{{\rm SR},k}}}),\label{eq116}\\
t_{10}&=\alpha^3 M_0M^2_1\beta^2_{{\rm SR},i}\beta_{{\rm RD},k}({\beta_{{\rm RD},i}+{M_1\beta_{{\rm RD},k}}}).\label{eq117}
\end{align}
Substituting (\ref{eq108}) into (\ref{eq117}), we have
\begin{align}\label{eq118}
{\mathbb{E}}\left\{{T_{k,i}T^H_{k,i}}\right\}&=\gamma^2\Big( t_1+t_2+t_3+t_4+2t_5+2t_6\notag\\
&+2t_7+2t_8+2t_9+2t_{10} \Big).
\end{align}
After some simplifications with $i=k$, we can obtain (\ref{eq38}).

\subsubsection{\rm{Compute $C_k$}}
Similar to the calculation of $B_k$, We can derive the expression (\ref{eq40}) of $C_k$.

\subsubsection{\rm{Compute $D_k$ and $E_k$}}
Similar to the calculation of $B_k$, we can obtain (\ref{eq41}) and (\ref{eq42}).

\subsubsection{\rm{Compute $F_k$}}
\begin{equation}\label{eq119}
\begin{aligned}
&{F_k}={\gamma ^2}{\mathbb{E}}\left\{ {\left|
{{\boldsymbol{\rm{g}}}_{{\rm{RD0}},k}^T{\boldsymbol {\rm G}}_{{\rm{RD0}}}^*{\mathbb{E}}\left\{ {{\boldsymbol {\rm G}}_{{\rm{SR1}}}^H\boldsymbol {\rm R}_{{\boldsymbol{\rm{n}}}_{{{\rm{q}}_{\rm{a}}}}{\boldsymbol{\rm{n}}}_{{{\rm{q}}_{\rm{a}}}}}{{{\boldsymbol {\rm G}}}_{{\rm{SR1}}}}}\right\}{{\boldsymbol {\rm G}}}_{{\rm{RD0}}}^T{\boldsymbol{\rm{g}}}_{{\rm{RD0}},k}^*}\right.} \right.\\
&+ \alpha {\boldsymbol{\rm{g}}}_{{\rm{RD0}},k}^T{\boldsymbol {\rm G}}_{{\rm{RD0}}}^*{\mathbb{E}}\left\{ {{\boldsymbol {\rm G}}_{{\rm{SR1}}}^H\boldsymbol {\rm R}_{{\boldsymbol{\rm{n}}}_{{{\rm{q}}_{\rm{a}}}}{\boldsymbol{\rm{n}}}_{{{\rm{q}}_{\rm{a}}}}}{{{\boldsymbol {\rm G}}}_{{\rm{SR1}}}}}\right\}{{\boldsymbol {\rm G}}}_{{\rm{RD1}}}^T{\boldsymbol{\rm{g}}}_{{\rm{RD1}},k}^*\\
&+ \alpha {\boldsymbol{\rm{g}}}_{{\rm{RD1}},k}^T{\boldsymbol {\rm G}}_{{\rm{RD1}}}^*{\mathbb{E}}\left\{ {{\boldsymbol {\rm G}}_{{\rm{SR1}}}^H\boldsymbol {\rm R}_{{\boldsymbol{\rm{n}}}_{{{\rm{q}}_{\rm{a}}}}{\boldsymbol{\rm{n}}}_{{{\rm{q}}_{\rm{a}}}}}{{{\boldsymbol {\rm G}}}_{{\rm{SR1}}}}}\right\}{{\boldsymbol {\rm G}}}_{{\rm{RD0}}}^T{\boldsymbol{\rm{g}}}_{{\rm{RD0}},k}^* \\
&+ \left. {\left. {{\alpha ^2}{\boldsymbol{\rm{g}}}_{{\rm{RD1}},k}^T{\boldsymbol {\rm G}}_{{\rm{RD1}}}^*{\mathbb{E}}\left\{ {{\boldsymbol {\rm G}}_{{\rm{SR1}}}^H\boldsymbol {\rm R}_{{\boldsymbol{\rm{n}}}_{{{\rm{q}}_{\rm{a}}}}{\boldsymbol{\rm{n}}}_{{{\rm{q}}_{\rm{a}}}}}{{{\boldsymbol {\rm G}}}_{{\rm{SR1}}}}}\right\}{{\boldsymbol {\rm G}}}_{{\rm{RD1}}}^T{\boldsymbol{\rm{g}}}_{{\rm{RD1}},k}^*} \right|} \right\}.
\end{aligned}
\end{equation}
Using results in (\ref{eq93}), we have
\begin{equation}\label{eq120}
{\mathbb{E}}\left\{ {{\boldsymbol {\rm G}}_{{\rm{SR1}}}^H\boldsymbol {\rm R}_{{\boldsymbol{\rm{n}}}_{{{\rm{q}}_{\rm{a}}}}{\boldsymbol{\rm{n}}}_{{{\rm{q}}_{\rm{a}}}}}{{{\boldsymbol {\rm G}}}_{{\rm{SR1}}}}}\right\}={\rm{diag}}\left({a_1,...,a_K}\right),
\end{equation}
where $a_n = M_1{\beta _{{\rm{SR}},n}}\left({\sum\limits_{i = 1}^K {{p_{{\rm S},i}}{\beta _{{\rm{SR}},i}}} +{p_{{\rm S},n}}{\beta _{{\rm{SR}},n}}+1} \right) .$
Substituting (\ref{eq120}) into (\ref{eq119}), we can obtain (\ref{eq42}).

\subsubsection{\rm{Compute $G_k$}}
With the help of (\ref{eq17}), we can obtain (\ref{eq43}) by applying similar approaches in the derivations of $\mathbb{E}\left\{{{\boldsymbol{\rm{R}}}_{{{{\boldsymbol{\rm{\tilde x}}}}_{{\rm{R0}}}}{{{\boldsymbol{\rm{\tilde x}}}}_{{\rm{R0}}}}}}\right\}$.

Combining all derived terms completes the proof.

\bibliographystyle{IEEEtran}
\bibliography{IEEEabrv,draft}

\begin{thebibliography}{10}
\providecommand{\url}[1]{#1}
\csname url@samestyle\endcsname
\providecommand{\newblock}{\relax}
\providecommand{\bibinfo}[2]{#2}
\providecommand{\BIBentrySTDinterwordspacing}{\spaceskip=0pt\relax}
\providecommand{\BIBentryALTinterwordstretchfactor}{4}
\providecommand{\BIBentryALTinterwordspacing}{\spaceskip=\fontdimen2\font plus
\BIBentryALTinterwordstretchfactor\fontdimen3\font minus
  \fontdimen4\font\relax}
\providecommand{\BIBforeignlanguage}[2]{{%
\expandafter\ifx\csname l@#1\endcsname\relax
\typeout{** WARNING: IEEEtran.bst: No hyphenation pattern has been}%
\typeout{** loaded for the language `#1'. Using the pattern for}%
\typeout{** the default language instead.}%
\else
\language=\csname l@#1\endcsname
\fi
#2}}
\providecommand{\BIBdecl}{\relax}
\BIBdecl

\bibitem{wong2017key}
V.~W. Wong, R.~Schober, D.~W.~K. Ng, and L.-C. Wang, \emph{Key Technologies for
  {5G} Wireless Systems}. Cambridge
  University Press, 2017.

\bibitem{marzetta2016fundamentals}
T.~L. Marzetta, E.~G. Larsson, H.~Yang, and H.~Q. Ngo, \emph{Fundamentals of
  Massive {MIMO}}. Cambridge University
  Press, 2016.

\bibitem{yadav2018all}
A.~Yadav and O.~A. Dobre, ``All technologies work together for good: {A} glance
  to future mobile networks,'' \emph{arXiv:1804.05963}, Apr. 2018.

\bibitem{morgado2018survey}
A.~Morgado, K.~M.~S. Huq, S.~Mumtaz, and J.~Rodriguez, ``A survey of {5G}
  technologies: {R}egulatory, standardization and industrial perspectives,''
  \emph{Digital Commun. Netw.}, vol.~4, no.~2, pp. 87--97, Apr. 2018.

\bibitem{zhang2017spectral}
J.~Zhang, X.~Xue, E.~Bj{\"o}rnson, B.~Ai, and S.~Jin, ``Spectral efficiency of
  multipair massive {MIMO} two-way relaying with hardware impairments,''
  \emph{IEEE Wireless Commun. Lett.}, vol.~7, no.~1, pp. 14--17, Feb. 2018.

\bibitem{wang2017performance}
Q.~Wang and Y.~Jing, ``Performance analysis and scaling law of {MRC/MRT}
  relaying with {CSI} error in multi-pair massive {MIMO} systems,'' \emph{IEEE
  Trans. Wireless Commun.}, vol.~16, no.~9, pp. 5882--5896, Sept. 2017.

\bibitem{jin2015ergodic}
S.~Jin, X.~Liang, K.~Wong, X.~Gao, and Q.~Zhu, ``Ergodic rate analysis for
  multipair massive {MIMO} two-way relay networks,'' \emph{IEEE Trans. Wireless
  Commun.}, vol.~14, no.~3, pp. 1480--1491, Mar. 2015.

\bibitem{bjornson2017massive}
E.~Bj{\"o}rnson, J.~Hoydis, and L.~Sanguinetti, ``Massive {MIMO} networks:
  {S}pectral, energy, and hardware efficiency,'' \emph{Foundations and
  Trends{\textregistered} in Signal Processing}, vol.~11, no. 3-4, pp.
  154--655, 2017.

\bibitem{zhang2018low}
J.~Zhang, L.~Dai, X.~Li, Y.~Liu, and L.~Hanzo, ``On low-resolution {ADC}s in
  practical {5G} millimeter-wave massive {MIMO} systems,'' \emph{IEEE Commun.
  Mag.}, vol.~56, no.~7, pp. 205--211, Jul. 2018.

\bibitem{Esswie2017novel}
A.~Esswie, M.~El-Absi, O.~A. Dobre, S.~Ikki, and T.~Kaiser, ``A novel {FDD}
  massive {MIMO} system based on downlink spatial channel estimation without
  {CSIT},'' in \emph{Proc. IEEE ICC}, May 2017, pp. 1--6.

\bibitem{li2010cooperative}
X.~Li, T.~Jiang, S.~Cui, J.~An, and Q.~Zhang, ``Cooperative communications
  based on rateless network coding in distributed {MIMO} systems,'' \emph{IEEE
  Wireless Commun.}, vol.~17, no.~3, pp. 60--67, Jun. 2010.

\bibitem{lee2008analog-to-digital}
H.~Lee and C.~G. Sodini, ``Analog-to-digital converters: {D}igitizing the
  analog world,'' \emph{Proc. IEEE}, vol.~96, no.~2, pp. 323--334, Feb. 2008.

\bibitem{zhang2016spectral}
J.~Zhang, L.~Dai, S.~Sun, and Z.~Wang, ``On the spectral efficiency of massive
  {MIMO} systems with low-resolution {ADC}s,'' \emph{IEEE Commun. Lett.},
  vol.~20, no.~5, pp. 842--845, May 2016.

\bibitem{fan2015uplink}
L.~Fan, S.~Jin, C.-K. Wen, and H.~Zhang, ``Uplink achievable rate for massive
  {MIMO} systems with low-resolution {ADC},'' \emph{IEEE Commun. Lett.},
  vol.~19, no.~12, pp. 2186--2189, Dec. 2015.

\bibitem{zhang2016achievable}
J.~Zhang, L.~Dai, X.~Zhang, E.~Bj\"{o}rnson, and Z.~Wang, ``Achievable rate of
  {R}ician large-scale {MIMO} channels with transceiver hardware impairments,''
  \emph{IEEE Trans. Veh. Techonol.}, vol.~65, no.~10, pp. 8800--8806, Oct.
  2016.

\bibitem{li2017channel}
Y.~Li, C.~Tao, G.~Secogranados, A.~Mezghani, A.~L. Swindlehurst, and L.~Liu,
  ``Channel estimation and performance analysis of one-bit massive {MIMO}
  systems,'' \emph{IEEE Trans. Signal Process.}, vol.~65, no.~15, pp.
  4075--4089, Aug. 2017.

\bibitem{liang2016mixed}
N.~Liang and W.~Zhang, ``Mixed-{ADC} massive {MIMO},'' \emph{IEEE J. Sel. Areas
  Commun.}, vol.~34, no.~4, pp. 983--997, Apr. 2016.

\bibitem{yuan2017distributed}
J.~Yuan, S.~Jin, C.-K. Wen, and K.~Wong, ``The distributed {MIMO} scenario:
  {C}an ideal {ADC}s be replaced by low-resolution {ADC}s?'' \emph{IEEE
  Wireless Commun. Lett.}, vol.~6, no.~4, pp. 470--473, Aug 2017.

\bibitem{liang2016frequency}
N.~Liang and W.~Zhang, ``Mixed-{ADC} massive {MIMO} uplink in
  frequency-selective channels,'' \emph{IEEE Trans. Commun.}, vol.~64, no.~11,
  pp. 4652--4666, Nov. 2016.

\bibitem{tan2016spectral}
W.~Tan, S.~Jin, C.~Wen, and Y.~Jing, ``Spectral efficiency of mixed-{ADC}
  receivers for massive {MIMO} systems,'' \emph{IEEE Access}, vol.~4, pp.
  7841--7846, 2016.

\bibitem{zhang2017performance}
J.~Zhang, L.~Dai, Z.~He, S.~Jin, and X.~Li, ``Performance analysis of
  mixed-{ADC} massive {MIMO} systems over {R}ician fading channels,''
  \emph{IEEE J. Sel. Areas Commun.}, vol.~35, no.~6, pp. 1327--1338, Jun. 2017.

\bibitem{zhang2016mixed}
T.~Zhang, C.~Wen, S.~Jin, and T.~Jiang, ``Mixed-{ADC} massive {MIMO} detectors:
  {P}erformance analysis and design optimization,'' \emph{IEEE Trans. Wireless
  Commun.}, vol.~15, no.~11, pp. 7738--7752, Nov. 2016.

\bibitem{pirzadeh2018spectral}
H.~Pirzadeh and A.~L. Swindlehurst, ``Spectral effciency of mixed-{ADC} massive
  {MIMO},'' \emph{IEEE Trans. Signal Process.}, vol.~66, no.~13, pp.
  3599--3613, Jul. 2018.

\bibitem{xu2017performance}
J.~Xu, W.~Xu, and F.~Gong, ``On performance of quantized transceiver in
  multiuser massive {MIMO} downlinks,'' \emph{IEEE Wireless Commun. Lett.},
  vol.~6, no.~5, pp. 562--565, Oct. 2017.

\bibitem{kong2017multipair}
C.~Kong, A.~Mezghani, C.~Zhong, A.~L. Swindlehurst, and Z.~Zhang, ``Multipair
  massive {MIMO} relaying systems with one-bit {ADC}s and {DAC}s,'' \emph{IEEE
  Trans. Signal Process.}, vol.~66, no.~11, pp. 2984--2997, Jun. 2018.

\bibitem{dong2017efficient}
P.~Dong, H.~Zhang, W.~Xu, and X.~You, ``Efficient low-resolution {ADC} relaying
  for multiuser massive {MIMO} system,'' \emph{IEEE Trans. Veh. Techonol.},
  vol.~66, no.~12, pp. 11\,039--11\,056, Dec. 2017.

\bibitem{liu2017multiuser}
J.~Liu, J.~Xu, W.~Xu, S.~Jin, and X.~Dong, ``Multiuser massive {MIMO} relaying
  with mixed-{ADC} receiver,'' \emph{IEEE Signal Process. Lett.}, vol.~24,
  no.~1, pp. 76--80, Jan. 2017.

\bibitem{kong2017full}
C.~Kong, C.~Zhong, S.~Jin, S.~Yang, H.~Lin, and Z.~Zhang, ``Full-duplex massive
  mimo relaying systems with low-resolution adcs,'' \emph{IEEE Trans. Wireless
  Commun.}, vol.~16, no.~8, pp. 5033--5047, Aug. 2017.

\bibitem{jia2016optimal}
X.~Jia, M.~Zhou, M.~Xie, L.~Yang, and H.~Zhu, ``Optimal design of secrecy
  massive mimo amplify-and-forward relaying systems with double-resolution adcs
  antenna array,'' \emph{IEEE Access}, vol.~4, pp. 8757--8774, 2016.

\bibitem{orhan2015low}
O.~Orhan, E.~Erkip, and S.~Rangan, ``Low power analog-to-digital conversion in
  millimeter wave systems: {I}mpact of resolution and bandwidth on
  performance,'' in \emph{Proc. ITA Workshop}, Feb. 2015, pp. 191--198.

\bibitem{fletcher2007robust}
A.~K. Fletcher, S.~Rangan, V.~K. Goyal, and K.~Ramchandran, ``Robust predictive
  quantization: Analysis and design via convex optimization,'' \emph{IEEE J.
  Sel. Topics Signal Process.}, vol.~1, no.~4, pp. 618--632, Dec. 2007.

\bibitem{max1960quantizing}
J.~Max, ``Quantizing for minimum distortion,'' \emph{IEEE Trans. Inf. Theory},
  vol.~6, no.~1, pp. 7--12, Mar. 1960.

\bibitem{liu2017pilot}
P.~Liu, S.~Jin, T.~Jiang, Q.~Zhang, and M.~Matthaiou, ``Pilot power allocation
  through user grouping in multi-cell massive {MIMO} systems,'' \emph{IEEE
  Trans. Commun.}, vol.~65, no.~4, pp. 1561--1574, Apr. 2017.

\bibitem{ngo2018total}
H.~Q. Ngo, L.-N. Tran, T.~Q. Duong, M.~Matthaiou, and E.~G. Larsson, ``On the
  total energy efficiency of cell-free massive {MIMO},'' \emph{IEEE Trans.
  Green Commun. Netw.}, vol.~2, no.~1, pp. 25--39, Mar. 2018.

\bibitem{boyd2004convex}
S.~Boyd and L.~Vandenberghe, \emph{Convex {O}ptimization}. Cambridge University Press, 2004.

\bibitem{he2014energy}
C.~He, B.~Sheng, P.~Zhu, D.~Wang, and X.~You, ``Energy efficiency comparison
  between distributed and co-located {MIMO} systems,'' \emph{Int. J. Commun.
  Syst.}, vol.~27, no.~1, pp. 81--94, Mar. 2014.

\bibitem{cui2005energy}
S.~Cui, A.~J. Goldsmith, and A.~Bahai, ``Energy-constrained modulation
  optimization,'' \emph{IEEE Trans. Wireless Commun.}, vol.~4, no.~5, pp.
  2349--2360, Sept. 2005.

\bibitem{lauwers2002power}
E.~Lauwers and G.~Gielen, ``Power estimation methods for analog circuits for
  architectural exploration of integrated systems,'' \emph{IEEE Trans. VLSI
  System}, vol.~10, no.~2, pp. 155--162, Apr. 2002.

\end{thebibliography}

\end{document}